\documentclass[a4paper,11pt]{article}
\pdfoutput=1 % if your are submitting a pdflatex (i.e. if you have
\usepackage{jheppub} % for details on the use of the package, please
\usepackage[T1]{fontenc} % if needed

\usepackage{chngpage}
\usepackage{tabularx}
\usepackage{tikz}
\usepackage{tikz-3dplot}
\usepackage{mathtools} 
\usepackage{extarrows} 
\usepackage{amsmath}
\usepackage{xcolor}
\usepackage{colortbl}

\definecolor{Gray}{gray}{0.85}
\definecolor{LightCyan}{rgb}{0.88,1,1}
\usetikzlibrary{positioning}
\usetikzlibrary{arrows}
\usetikzlibrary{decorations.pathreplacing}
\usetikzlibrary{shapes.geometric}
\usepackage{mathtools}
\usepackage{wasysym} 

\usetikzlibrary{knots}
\usepackage[english]{babel}
\usepackage{amsthm}
\theoremstyle{definition}

\def\a {\alpha}
\def\b{\beta}
\def\F{\text{F}}
\def\eff{\text{eff}}
\def\F{\textbf{F}}
\def\AF{\textbf{AF}}
\def\inf{\infty}
\def\bsp{\begin{split}}
\def\esp{\begin{split}} 

\def\N{\mathcal{N}}

\def\R{\mathbb{R}}
\def\M{\text{M}}
\def\D{\text{D}}
\def\NS{\text{NS}}
\def\Z{\mathbb{Z}}
\def\CS{\text{Chern-Simons }}

\def\A{\text{A}}
\def\B{\text{B}}

\def\redK{{\color{red}K}}
\def\redS{{\color{red}S^1}} 
\def\N{\mathcal{N}}

\def \redcirc{{\color{red}{\bigcirc}}}
\def\bluecirc{{\color{blue}{\bigcirc}}}
\def\redK{{\color{red}K}}
\def\graybox{ {\color{gray}\blacksquare} }
\def\tL{\widetilde{L}}
\def\IIA{\text{IIA}}
\def\IIB{\text{IIB}}
\def\Ima{\text{Im}(\alpha)}
\def\Imb{\text{Im}(\beta)}
\def\fa{f(\alpha)}
\def\fb{f(\beta)}

\def\L{\mathcal{L}}

\tikzset{gauge/.style={circle,fill,inner sep=2pt}}
\tikzset{matter/.style={fill=gray, inner sep=2.4pt,regular polygon,regular polygon sides=4}}

\title{\boldmath 
Kirby Moves for 3d Gauge Theories
}

\author[a,b,c]{Shi Cheng}
\affiliation[a]{Center for Mathematics
and Interdisciplinary Sciences, Fudan University
Shanghai 200433, China}

\affiliation[b]{Shanghai Institute for Mathematics
and Interdisciplinary Sciences
Shanghai 200433, China}

\affiliation[c]{Department of Physics and Center for Field Theory \& Particle Physics, Fudan University, 20005, Songhu Road, 200438 Shanghai, China}

\emailAdd{shicheng@simis.cn}

\abstract{The geometric engineering of 3d $\mathcal{N}=2$ gauge theories can be given by M5-branes on three-manifolds. We focus on the Dehn surgeries and Kirby moves of three-manifolds. We notice the Kirby move can be slightly extended when the chiral multiplets exist, and these chiral multiplets arise from Lagrangian M5-branes of the Ooguri-Vafa type in the cotangent bundles of three-manifolds. Through string dualities, we check that this M2-M5-brane configuration is consistent with 3d brane webs. We start from analyzing lens spaces as a toy model, and extend results to plumbed three-manifolds. We also show a basic 3d duality can be interpreted as Rolfsen twist. 

}

\begin{document} 
\maketitle
\flushbottom

\section{Introduction}
String theory leads to many interesting results of 3d $\N=2$ gauge theories \cite{Intriligator:1996ex,Aharony:1997aa,Hanany:1996ie,Kapustin:1999ha,Dorey:1999rb,Giveon_2009,Benini:2011aa,Intriligator:2013lca,Alday:2017yxk} and provides constructions including  geometric engineering and brane webs \cite{Boer:1997ts,Benvenuti:2016wet,Cheng:2021vtq}. The difficulty of the construction lie on mixed Chern-Simons levels and superpotentials.  Fortunately, three-manifold as a relative new introduced geometry to gauge theories is promising  due to their fruitful mathematical structures, such as Kirby moves and JSJ decomposition \cite{kirbymove}. In this paper, we focus on the equivalent transformations of the three-manifolds. These structures have been entered the visionary story of  3d-3d correspondence \cite{Dimofte:2011ju,Gukov:2016ad,Cecotti:2011iy,Terashima:2011qi}.

The M5-branes should wrap three-manifolds, and experience the  surgeries of three-manifolds. The gauge theory on M5-branes are 6d $(2,0)$ superconformal field theories, and after the compactification  on closed three-manifolds, the 3d gauge theories are created and  labeled by the three-manifolds. For example, in the case of lens spaces $L(k,1)$, one can get 3d $\mathcal{N}=2$ gauge theories $U(1)_k$. In this paper, we consider more generic three-manifolds, called
 plumbed manifolds.

\begin{figure}[h!]
	\centering
	%	\begin{split}
		\includegraphics[width=2.5in]{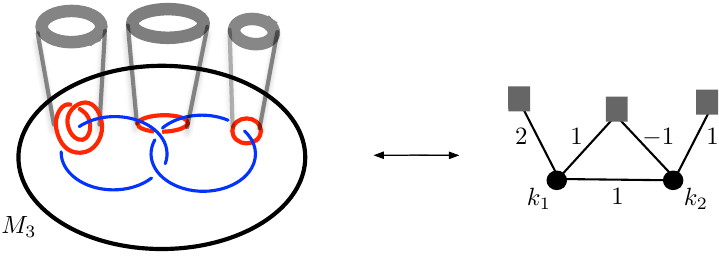}
		\caption{Blue circles $\bluecirc$ denote surgical circles. Red circles $\redcirc$ are matter circles denoting intersections and are boundaries of M2-branes. This combined three-manifold corresponds to a 3d theory  with two gauge groups $U(1)_{k_1} \times U(1)_{k_2}$, and three chiral multiplets $\Phi_i$ with charges $(q_{i}^1, q_{i}^2)$.
On the right, we represent it by a plumbing graph.}
		\label{fig:ex0matter}
		%	\end{split}
\end{figure}

The motivation of this paper is simple.
 The abelian 3d gauge theories should have mixed Chern-Simons levels given by the linking numbers of plumbed manifolds \cite{Gadde:2013aa}. 
 In \cite{Cheng:2023ocj}, this interpretation is extended by including chiral multiplets. 
 Then the question is how  to engineer these chiral multiplets? Perviously, there is no clear answers. 
 To describe chiral multiplets, it is convenient to introduce some gray boxes to denote chiral multiplets and decorate plumbing graphs, as shown in Figure \ref{fig:ex0matter}. From the perspective of gauge theories, it is fine to do this extension, because gauge theories usually contain chiral multiplets, and the decorated plumbing graphs work well when discussing gauge theory dualities. We are targeting to translate these decorated plumbed graphs to three-manifolds. We should consider many things, such as Kirby moves of various types and the so on, to ensure geometrically these chiral multiplets do not break the Kirby move of three-manifolds. But the first thing to find is what geometric object  engineer the chiral multiplets. We can use the M2-M5-brane configurations and string dualities to analyze. It turns out that these chiral multiplets should be the open strings connecting the closed three-manifolds and the Lagrangian branes of the Ooguri-Vafa (OV) type \cite{Ooguri:1999bv}. 
Then we need to introduce these OV Lagrangian branes in the cotangent bundles of the three-manifolds.

Now we get a new geometric object as the OV Lagrangian branes, and we should  know how they interact with the three-manifolds. By studying the Dehn surgeries, we note that the OV Lagrangian branes should link to the surgical knots that label the three-manifolds, and the linking number is the charge of the chiral multiplet.
This is not a surprise, since in the geometric engineering using Calabi-Yau manifolds, charges of chiral multiplets are often realized as the intersection numbers between divisors and curves. 
 The next thing is to geometrically understand a basic physical duality involving the chiral multiplets. 
Using an idea inspired by identical surgery, we propose that this basic duality should be interpreted as the Rolfsen twist of the three-manifold. This twist is special and not falls in the scope of the Kirby moves, but does lead to equivalent surgeries.

 Let us briefly review the Ooguri-Vafa construction to show the results of this paper. To begin with, M5-branes wrapping the Lagrangian submanifold  intersect the M5-branes wrapping the three-sphere along a knot: $L_\redK \cap S^3=\redK$. The chiral multiplet only corresponds to the unknot, and the three-sphere can be replaced by generic three-manifolds, and hence  $L_{\color{red}\bigcirc} \cap M_3 = \redcirc$ where we use the red circle to  distinguish it from the surgical knots and links. 
The main tool that we will use  is the M-theory/IIB string duality, which maps the Ooguri-Vafa M5-branes to D5-branes that engineer flavor symmetries of the open strings. The D3-branes come from the M5-branes wrapping the base three-manifold $M_3$. There are many freedoms to intersect the Ooguri-Vafa M5-brane with the three-manifold $M_3$. The unknot $\redcirc$ could link to any components of the surgical links  and hence the chiral multiplets could be charged under abelian gauge groups $U(1) \times U(1) \times \cdots U(1)$. 

\begin{figure}[h!]
	\centering
	\includegraphics[width=1.5in]{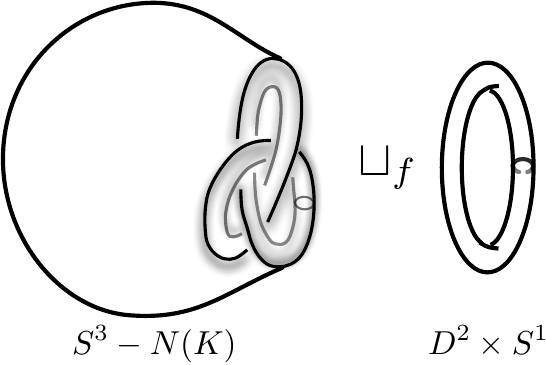}
	\qquad\qquad \includegraphics[width=1.82in]{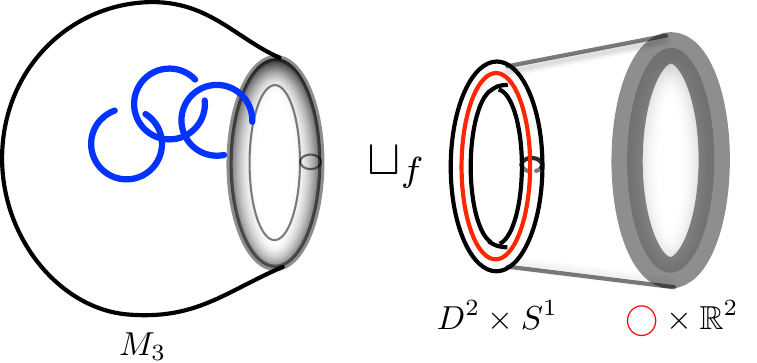}
	\caption{The left figure illustrates the three-manifold is obtained by the surgery over a trefoil. The right figure shows how the Lagrangian brane of the OV type is linked to the surgical links colored in blue. }
	\label{fig:surgeryKde}
\end{figure}

This surgery construction is illustrated in Figure \ref{fig:surgeryKde} and the dictionary is shown in Table \ref{tab:dic}. In Figure \ref{fig:surgeryKde}, the left graph is the Dehn surgery of three-manifolds $M_3$, defined by drilling out the neighborhood of a knot $K$, and then filling in a solid torus. The right graph shows the Ooguri-Vafa Lagrangian $L_\redcirc$ is introduced by attached it to a solid torus.
Let us use plumbing graphs to represent 3d theories, and show an example in Figure \ref{fig:ex0matter}.
Note that non-compact Ooguri-Vafa Lagrangian branes are located in the cotangent bundle of the three-manifold $T^*M_3$, and hence are parallel.
\begin{table}%[h!]
\centering
		\text{
			\begin{tabular}{ c| c | c  }
				\hline
				{three-manifolds}  & plumbing graphs & { abelian gauge theories}	\\ \hline
				Lagrangian submanifold $\color{red} \bigcirc$& $\color{gray}\blacksquare$~\, &  chiral multiplet \\ 
				surgery circle $\color{blue} \bigcirc$& $ {\bullet}_{k_i}$ &  gauge group $U(1)_{k_i}$\\ 
				linking number between $\color{red} \bigcirc$ and $\color{blue} \bigcirc$&
			$			\begin{tikzpicture}
				\draw[thick] (0,-0.2) --(1,-0.2)node[midway,above=-0.1]{\tiny$q_i$};	
				\node at (0.0,-0.2) {$\bullet$}  ;
				\node at (1,-0.2) {$\graybox$};
				\end{tikzpicture}
				$
				%$\bullet_k \ \graybox$ 	 num. on lines between $\color{gray}\blacksquare$ and $\bullet$
				 & charge $q_i$ \\
				 
				linking number between $\color{blue} \bigcirc$ and $\color{blue} \bigcirc$	& %num. on lines between two $\bullet$s 
					$\begin{tikzpicture}
					\draw[thick] (0,0) --(1,0)node[midway,above=-0.1]{\tiny$k_{ij}$};	
					\node at (0.0,-0) {$\bullet$}  ;
					\node at (1,0) {$\bullet$} ;
				\end{tikzpicture}$
				 & effective CS levels $k_{ij}$  \\
				Kirby moves on  $\color{blue} \bigcirc$  & blow up/down $\bullet_k$& integrate in/out $U(1)_k$ \\
				
	Rolfsen twist \eqref{effST}
				&$ST$-move  \eqref{plumbSTk} & gauged mirror triality \\
				
					%handle-slides of $\color{red} \bigcirc $ \eqref{handleslidescircle}
				%& 
				%	add $\graybox$
					 % & add chiral multiplets  \\
			%	Dehn twists& effective framing numbers  & effective CS levels \\
				\hline
			\end{tabular}		}
		\caption{The dictionary between three-manifolds, plumbing graphs, and gauge theories. We often refer to the red circle $\redcirc$ as the matter circle.}
		\label{tab:dic}
\end{table}
We clarify that the red circle and the blue circle form a Hopf link in the solid torus:
 \begin{align}	
 \label{mgcircles}
	\centering
	\begin{split}
		\includegraphics[width=0.6in]{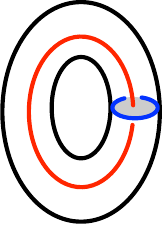}
	\end{split}
\end{align}
The red circle locates at the core of the solid torus, which is often called the singular fiber. Note that it is not the longitude on the boundary of the solid torus. The blue circle corresponds to the gauge group and is the meridian of the boundary. The red circle and the blue circle cannot pass through each other, so they form a Hopf link.

The outline is as follows.
In Section \ref{sec:charges}, we discuss a gauge theory duality of plumbing graphs and explain it as a new Kirby move.
In Section \ref{sec:matterdefect},  through string duality and brane webs, we show that Lagrangian M5-branes engineer chiral multiplets.
In Section \ref{sec:branewebs}, we discuss brane webs to compare with the M-theory configurations on lens spaces.
In Section \ref{sec:matterinthree}, we extend the results to plumbed three-manifolds, and show the special properties of gluing maps transport to 3d brane webs. In Section \ref{sec:STdrill}, we proposed that the new Kirby move is actually the Rolfsen twist, inspired by identical surgery.

\section{A new Kirby move}\label{sec:charges}

Kirby moves describe equivalent surgeries for the same three-manifold. For analyzing pure gauge theories without chiral multiplets, Kirby moves are already enough, but if chiral multiplets exist, the standard Kirby moves in the textbook are not sufficient and should be extended. In this section, we show  a basic gauge theory duality that can be translated as a new Kirby move. 

\subsection{ST-move}
The basic duality is between the free chiral multiplet and the $U(1)_{\pm 1/2}$ theory with a charge one fundamental chiral multiplet:
\begin{equation}\label{basicmirror}
	1 \F ~~\longleftrightarrow ~~ U(1)_{\pm \frac{1}{2} } +1\F  \,,
	\end{equation}
which can be represented by  the plumbing graph:
\begin{equation}\label{mirrortriality0}
		\begin{aligned}
			\includegraphics[width=1in]{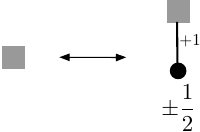}
		\end{aligned}
	\end{equation}
	where $\pm1/2$ are bare \CS levels.
This duality is noticed in \cite{Dimofte:2011ju} by decoupling the anti-fundamental matter $1\AF$ of the 3d $\N=4$ mirror pair:
	\begin{align}
	1\F +1\AF~~\longleftrightarrow~~	U(1)_0 + 1\F +1\AF  \,.
	\end{align}
	In \cite{Cheng:2020aa,Cheng:2023ocj} , we noticed that if  the dual $U(1)$ global symmetry  on both sides is gauged, a new dual pair is generated:
\begin{equation}\label{bareSTq}
\begin{aligned}
	\includegraphics[width=1.4in]{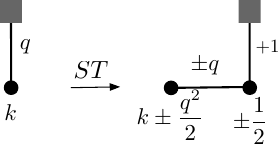}
	\end{aligned}\,,
	\end{equation}
	where $q$ is the charge of the fundamental chiral multiplet.
    Moreover, this dual pair \eqref{basicmirror} corresponds to $ST \in SL(2,\mathbb{Z})$ in \cite{Witten:2003ya},  so \eqref{bareSTq} is named $ST$-move in this paper. One can also assign effective Chern-Simons couplings to the plumbing graphs, which receive quantum corrections from chiral multiplets \cite{Intriligator:1996ex}:
	 \begin{equation}\label{effKij}
	 	k^{\eff} = k^{\text{bare}}+ \sum_{I=1}^{N_f}\frac{q^2}{2} \, \text{sign}(q_I) \text{sign}(m_I)  \,.
	 	\end{equation}
Using the effective Chern-Simons couplings, we can write the ST-move as
    \begin{equation}\label{effSTm}
\begin{aligned}
	\includegraphics[width=1.7in]{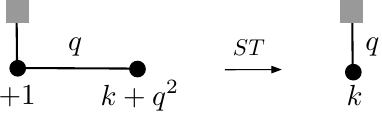}
\end{aligned}
\end{equation}
    In the case of the chiral multiplet coupled to many gauge nodes, one can still apply the ST-move. The chiral multiplets of this type are multiple charged. We illustrate a bifundamental chiral multiplet with charges $(q_1,q_2)$ and its ST-move as follows:
 \begin{equation}\label{bibeta}
	\raisebox{-.5\height}{
		\begin{tikzpicture}[scale=0.8]
			\node at (7,0){
				$ \begin{tikzpicture}[scale=0.8]
					\node[gauge,label=below:{$+1$}] (1) at (0,0) {};
					\node[matter] (2) at (0,0.6) {};
					\draw[thick] (1)--(2);
					\node[gauge,label=right:{$k_1+q_1^2$}] (3) at (1.5,1) {};
					\node[gauge,label=right:{$k_2+q_2^2$}] (4) at (1.5,-1) {};
					\draw[thick,red]  (3)--(4) node[midway, right]{$k+q_1q_2$} ; 
					\draw[thick](1)--(3)node[midway,above]{$q_1$}
					(1)--(4)node[midway,below]{$q_2$};
				\end{tikzpicture}
				$  } ;
			\draw[thick,-latex] (9.5,-0.2)--(10.6,-0.2)	;
			\node at (10,0.1) {\small${ST}$};
			\node at (12,0){
				$ \begin{tikzpicture}[scale=0.8]
					\node[matter](1) at (0.6,0){} ;
					\node[gauge,label=right:{$k_1$}] (3) at (1.5,1) {};
					\node[gauge,label=right:{$k_2$}] (4) at (1.5,-1) {};
					\draw[thick,red]  (3)--(4) node[midway, right]{$k$} ; 
					\draw[thick]   (1)--(3)node[midway, above]{$q_1$}
					(1)--(4)node[midway, below]{$q_2$}
					;
				\end{tikzpicture}
				$
			}
			;
		\end{tikzpicture}
	}
\end{equation}
For more details of computation in terms of partition functions, see \cite{Cheng:2021nex}.
 This new dual pair gives an extension of Kirby moves, in the sense that the gray box is added to the plumbing graphs. It is easy to see that if the chiral multiplets (gray boxes) on both sides are decoupled or deleted, ST-move reduces to the blow-up or blow-down as  the first type of Kirby moves \cite{kirbymove}.
 
 The above examples such as \eqref{bibeta} indicate that the gray box should correspond to some circles in three-manifolds because the charge $q_i$ can be interpreted as the linking numbers of  plumbing graphs representing surgeries. In the next section, we will argue that the gray boxes come from some geometric objects, which is a  main discovery of this work. Readers can jump to the last section \ref{sec:STdrill}  to see the geometric derivation of this extended Kirby move.
    
\subsection{Handle slides}\label{sec:handleslides}
The Kirby moves that we have just shown are the first type, and there is  a second type called handle slides. 
The blow-up and blow-down have already been given a physical interpretation in \cite{Gadde:2013aa,Cheng:2023ocj}. In particular, for abelian theories, this first type is integrating in or out gauge fields. The second type is not extensively studied previously, and we show that it is also useful and even leads us to  the geometric realization of chiral multiplets. 

Let us  mention the definition of the handle slides, which are linear combinations of surgerical circles and can be illustrated as
\begin{equation}\label{betaKirbyex01}
	\begin{aligned}	\includegraphics[width=3in]{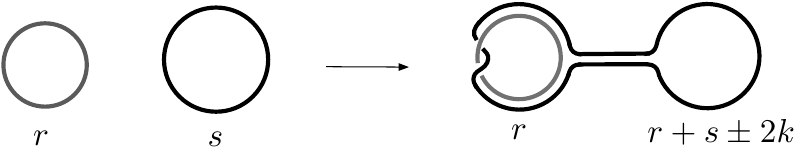}
	\end{aligned}
\end{equation}
Let us use $L_1$ and $L_2$ to denote two surgical circles with linking numbers:\begin{equation}\label{oldlink}
	L_1 \cdot L_1 =r\,,~~  L_1 \cdot L_2=   L_2 \cdot L_1 =k\,,~~L_2\cdot L_2=s\,,
\end{equation}
What is interesting is that after the handle slide, one circle $L_1$ is not affected, while the other circle becomes a connected sum $L_1+L_2$, and these two  circles after sliding are linked.
More explicitly,
$L_1$ becomes $\tilde{L}_1=L_1$, and $L_2$ becomes $
\tilde{L}_2 =  L_2  \pm L_1 
$. The linking numbers become
\begin{equation}
	\tL_1\cdot \tL_1=r\,,~~  
	\tL_1 \cdot {\tL}_2={\tL}_2 \cdot \tL_1 =k \pm r\,,~~
	\tilde{L}_2 \cdot {\tL}_2 =r+s\pm 2k 
	\,.
\end{equation}

From the physical perspective, the handle slides also give equivalent theories, since integrating the gauge node $U(1)_r$ marked by orange circles in the following leads to the same theory:
\begin{equation}\label{integrL1}
	\begin{aligned}
		\includegraphics[width=3in]{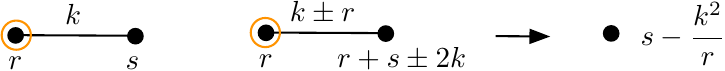}
	\end{aligned}\,.
\end{equation}
In general, one can recursively apply handle slides. Then any $\tilde{L}_2=L_2 + n L_1$ with $n\in \mathbb{Z}$ are equivalent to the node $U(1)_{s-\frac{k^2}{r}}$.  For generic $n$, the handle slide is
\begin{equation}\label{nbetamove}
	\begin{aligned}
		\includegraphics[width=2.2in]{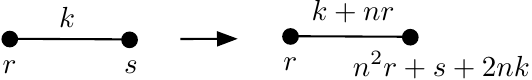}
	\end{aligned} \,.
\end{equation}
Handle slides can be reversed, as $n$ can be negative integers. 
There is a little overlap between blow-up/down and handle slides, which happens in the case $r=1$.
In our following work \cite{Cheng:2024ybd}, we extensively use handle slides as a key tool to simplify the gauge theories and various identities.

\vspace{4mm}
\noindent
\textbf{A puzzle on framing numbers.} 
Framing numbers are self-linking numbers, or in other words, the Chern-Simons levels of the gauge nodes of plumbing graphs. In the following, we mention a puzzle that one should be careful about.

It is better to view the linking numbers of plumbing graphs as effective Chern-Simons levels of abelian 3d theories with many $U(1)$ gauge groups.  
We should only consider plumbing graphs with integral framing numbers to avoid parity anomalies of 3d theories, and if fractional framing numbers are present, they should be resolved through Kirby moves. 

Even through, bare \CS levels can be assigned on plumbing graphs, but there are subtleties involved.
 If using bare CS levels, one should add quantum corrections from chiral multiplets to framing numbers.
 For example, the $ST$-move shown in \eqref{bareSTq} can be applied recursively to a chiral multiplet and gives a chain $\bullet_{k\pm q^2 / 2}-\bullet_{\pm1}-\cdots-\bullet_{\pm1}-\bullet_{\pm1/2}-\graybox$, which is absolutely correct, and any $\bullet_{\pm1}$ in the middle can be integrated out.
However, if one applies the ST-move many times using effective levels in \eqref{effSTm}, one gets an over-corrected tail $-\bullet_2-\bullet_2-\bullet_1-\graybox$, which is wrong, since integrating out middle gauge nodes $\bullet_{2}$ could lead to a wrong result that conflicts with the ST-duality. The correct one should be $-\bullet_{\pm1}-\bullet_{\pm1}-\bullet_{\pm1}-\graybox$. For more details, see examples in \cite{Cheng:2023ocj}. Surely, if the chiral multiplet is not present, the tail $-\bullet_2-\bullet_2-\bullet_1$ is absolutely correct, since it can be obtained by blow-up. This example shows the crucial difference between the new Kirby move (ST-moves) and the standard Kirby move of the blow-up. This difference lies in the fact that the quantum correction from the chiral multiplet is a half-integer. To avoid this subtle problem, we do not apply ST-moves recursively and only apply the ST-move at most one time, while the gauge nodes without chiral multiplets can be applied Kirby moves recursively.

\section{ M-theory configuration}\label{sec:matterdefect}
 If we wrap M5-branes on three-manifolds, 3d theories can be obtained and denoted by $T[M_3]$, since the 3d theories are determined by three-manifolds $M_3$.
In this section, we  consider lens spaces as a toy model, since lens spaces have dual 3d brane webs in type IIB string theory, and  the 3d brane webs should carry the corresponding properties of membrane configurations on lens spaces.
From this toy model, we notice that the M5-branes wrapping the Ooguri-Vafa Lagrangian brane are dual to the D5-branes in  3d brane webs, and hence engineer flavor symmetries. Finding the right directions to put branes and perform the string duality is the technical difficulty that we will address. 

\subsection{Lens spaces and brane webs}\label{sec:defectm5}
\vspace{4mm}\noindent
\textbf{Spacetime.}
To begin with, lens spaces $L(k,1)$ are embodied in the 11-dimensional spacetime of the M-theory:
\begin{align}
 \R_\epsilon^2\times S^1 \times  T^*L(k,1) \times \R^2  \,,
 \end{align}
where the 3d theories $T[L(k,1)]$ live on the spacetime $\mathbb{R}^2 \times S^1$, and $T^*L(k,1)$ as the cotangent bundle of the lens space, whose fiber as the normal direction is $\R^3$. The R-symmetry $U(1)_R$ is the rotation symmetry of the last term $\R^2$.

The 11-th direction of M-theory is denoted by $S^1_\sharp$, arising from the strong coupling limit of type IIA string theory. If reducing the M-theory to type IIA string theory, this direction will shrink. This M-theory circle $S_\sharp^1$ should take one direction in $T^\star L(k,q)$, and at first sight there are more than one candidates, so it is kind of non-trivial to identify this circle. Fortunately, through the dual 3d brane web we can identify this  circle should be in the lens space: $S^1_\sharp  \subset  L(k,1)$. To begin with, the lens spaces are elliptically fibered over an interval $T^2 \hookrightarrow L(k,1) \rightarrow I$. Equivalently, it can be viewed as gluing two solid tori $D^2 \times S^1$ along their boundaries, namely a Heegaard splitting of the lens space, and the meridian $S^1 =\partial (D^2)$ of the solid torus shrinks at the two endpoints of the interval $I$. This meridian is actually the $S^1_\sharp$ that we will verify soon.

Moreover,
theories with a gauge group $U(N_c)_k$ are given by wrapping $N_c$ number of M5-branes on $\R_\epsilon^2\times S^1 \times L(k,1)$. In this paper, we set $N_c=1$ and only consider a single M5-brane, hence the 3d theories obtained should be abelian. If considering more generic three-manifolds such as plumbed three-manifolds, the gauge groups would contain many $U(1)$. 
These manifolds are often non-hyperbolic three-manifolds, hence are somehow different from the manifolds used in the DGG construction \cite{Dimofte:2011ju}. We do not care whether these manifolds have hyperbolic metrics or not at this moment, since the metrics of the three-manifolds are not very relevant.

\vspace{4mm}\noindent
\textbf{Ooguri-Vafa construction.}
This construction is proposed in \cite{Ooguri:1999bv}, providing a string theory construction for 3d Chern-Simons theories with Wilson loops as well as open topological strings. If putting this construction in M-theory, it involves M2-branes and M5-branes. These M2-branes give open topological strings and should end on M5-branes wrapping the three-manifolds. The Ooguri-Vafa (OV) Lagrangian M5-brane locates in the contangent bundle $\L_K \subset T^*S^3$, where $K$ is a knot and denotes the intersection between three-manifolds $K= \L_K \cap S^3$.

This Lagrangian submanifold $\L_K$ is non-compact, and hence gives a flavor symmetry $U(1)_F$. The base three-sphere $S^3$ is wrapped by another  M5-brane, and gives a gauge group $U(1)_k$ where the Chern-Simons level is given by the framing number of the lens space. In particular, this three-sphere can be assigned with many equivalent framing number $1/n$ with $n \in \Z$, while only $\pm1$ and $1/0$ looks meaningful as the Chern-Simons levels can only be integers. 
In addition,
the matter fields are hypermultiplets charged under $U(1)_G\times U(1)_F$, and come from the M2-branes stretching between $S^3$ and $\L_L$.

\vspace{4mm}
\noindent
\textbf{String dualities.}
 The string duality between M-theory and IIB string provides a method to  check the consistency of the construction.  For clarity,  we assign numbers to spacetime as follows:
\begin{equation}\R_{12}^2\times S_0^1 \times N_{345} \times  L(k,1)_{69\sharp} \times \R^2_{78}  \,.
	\end{equation} 
The elliptic fiber bundle reads $ T^2_{9\sharp} \hookrightarrow L(k,1) \rightarrow I_6$, where the fiber is a torus $T^2_{9\sharp}=S^1_9\times S^1_\sharp $ with longitude  $S^1_9$  and meridian  $S^1_\sharp$.  Distinguishing  the meridian and longitude is important, since they cannot be switched freely, and even in IIB string theory, S-duality is  not preserved for generic 3d $\N=2$ theories. 

The M-theory/IIB duality is given by a T-duality and a S-duality.
If the T-duality is applied on a direction of $\D p$-branes,  the brane shrinks along this direction and reduces to a $\D(p-1)$-brane:
\begin{align}
\text{Dp wraps on }x_i ~~\xlongleftrightarrow{\text{T-dual along } x_i }~~\text{D(p-1) at a point on }x_i \,.
\end{align}
The S-duality in type IIB string theory exchanges D5-branes and NS5-branes,  F1-strings and D1-branes, but preserves D3-branes.
The duality between M-theory and type IIB string theory can be represented as $\M/T_{9\sharp}^2 \simeq \IIB/S_9^1$ as follows:
\begin{align}
	\text{M-theory} ~\xrightarrow{ ~~\text{shrink}~ S^{1}_{\sharp}~}~ \text{IIA}~ \xrightarrow{~ \text{T-dual along }x_9~ }~\text{IIB} \,.
\end{align}
Here we have chosen a direction for the T-duality.
Through this duality, the M5-branes wrapping the lens spaces become 3d brane webs that contain  D3-branes, D5-branes and NS5-branes. M2-branes become the F1-strings and D1-branes on the 3d brane webs. The 3d brane webs provide symmetry groups, and  F1/D1-strings lead to chiral multiplets.

\vspace{4mm}\noindent
\textbf{The dictionary.}
 In the above, we have guessed that the Ooguri-Vafa Lagrangian brane should engineer the chiral multiplet, and here we provide the evidence.

\begin{table}%[h!]
    \centering
    \begin{tabular}{|c|c|c c c| c  c | c | c | c c  | c  c |  }
   \multicolumn{1}{c}{} &   \multicolumn{1}{c}{} & \multicolumn{3}{c}{ $\overbracket{\qquad\qquad}^{S^1 \times \R^2}$}  &
     \multicolumn{3}{c}{$\overbracket{\qquad\qquad}^{N_{345}}$ } &
     \multicolumn{3}{c}{$\overbracket{\qquad\qquad}^{I_6 \times T^2_{9\sharp} }$ } &
       \multicolumn{2}{c}{$\overbracket{\qquad}^{ \R^2 } $}
    \\ \hline
 %   \multicolumn{3}{c |c c}{Country}  \\
  %  & $\R_{\epsilon}^2 \times S^1$ &  \\
\text{11d} & branes & 0 & 1 & 2 & 3 & 4 &  {\colorbox{LightCyan}5} & 6 & {\colorbox{LightCyan}9}  &$\sharp$ & 7 & 8 \\ \hline
     \rowcolor{Gray}
M-theory  &     $N_c$ M5  & 0 & 1 & 2 &  &  &  & 6 & $9_\A$ &$\sharp$ &  &  \\
IIA &       $N_c$ D4  & 0 & 1 & 2 &  &  &  & 6 & $9_\A$ & &  &  \\
      \rowcolor{LightCyan}
IIB &  $N_c$ {\color{blue}D3}   & 0 & 1 & 2 &  &  &  & 6 &  & &  &  \\ \hline
%IIA &  D0   &  &  &  &  &  &  &  &  & $\sharp$ &  &  \\
IIA &    D6   & 0 & 1 & 2 & 3 & 4 & 5 &  & $9_\A$ & &  &  \\
    \rowcolor{LightCyan}
IIB &  $\D5 \xrightarrow{\scriptstyle S} {\color{blue}\NS5}  $ & 0 & 1 & 2 & 3 & 4 & 5 &  &  & &  &  \\
\hline
   \rowcolor{Gray}
M-theory &   $\M5''$  & 0 & 1 & 2  & 3  & 4 &  &  & ${\color{red}9_\A}$ &   &  &  \\
IIA &   $\NS5''$  & 0 & 1 & 2  & 3  & 4 &  &  & $9_\A$ &   &  &  \\
   \rowcolor{LightCyan}
IIB &   $\NS5''\xrightarrow{ S} {\color{red}\D5}  $ & 0 & 1 & 2  & 3  & 4 &  &  & $9_\B$ &   &  &  \\
M-theory &    $\M2$ & 0 &  &  &   &  & 5 &  &  $9_\A$ &  &  & \\
IIB &    $\D1 \xrightarrow{{S}} \text{F}1$ & 0 &  &  &   &  & 5 &  &  &  &  &  \\ \hline\hline
   \rowcolor{Gray}
M-theory&   $\M5'$  & 0 & 1 & 2 & 3 & 4 &   &  &  &$\sharp$ &  &  \\
IIA &   $\D4'$  & 0 & 1 & 2 & 3 & 4 &   &  &  & &  &  \\
IIB &   $\D5'\xrightarrow{ S} \NS5 $  & 0 & 1 & 2 &  3 & 4 &  &  & $9_\B$ & &  &  \\ 
M-theory &   $\M2$ & 0 &  &  &   &  & 5 &  &  &  $\sharp$ &  & \\
IIB &   $\text{F}1 \xrightarrow{{S}} \text{D}1$ & 0 &  &  &   &  & 5 &  &  &  &  &  \\
\hline
    \end{tabular}
    \caption{This table shows various branes  allowed at one endpoint of the interval $I_6$. Since the direction $x_9$ appears in both IIA and IIB, we use a subscript in $9_\A$ and $9_\B$ to distinguish them. We use blue and red colors to emphasize some branes.
    The branes on the other endpoint of $I_6$ can be obtained similarly. 
    }
    \label{tab:branedefts}
\end{table}

The Ooguri-Vafa Lagrangian brane $L_\bigcirc$ has various ways to intersect the lens spaces, but we cannot expect that all choices are meaningful. After tried all choices, we find a canonical one, which ensures that M-theroy configuration is dual to a standard 3d brane web. We  summarize the details in Table \ref{tab:branedefts}. Let us explain how to obtain them.
To begin with, the gauge group is engineered by a single M5-brane wrapping the lens space $L(k,1)$. This M5-brane reduces to a D4-brane after shrinking $S^1_\sharp$, and then the T-duality along $S^1_9$ of the torus $T^2_{9\sharp}$ turns it into a D3-brane which engineers a gauge group $U(1)_k$. In short, $\M5(1269\sharp) \rightarrow \D4 (1269) \rightarrow \D3(126)$. This process is presented in the table. The string duality that we applied is
\begin{align}
\text{M-theory} \xrightarrow{/ S^1_\sharp} \text{IIA} \xrightarrow{\text{T}_9} \text{IIB} \xrightarrow{\text{S}} \text{IIB} \,.
\end{align}

The bare \CS level of this gauge group $U(1)_k$ is related to the relative angle $\theta$ between two NS5-branes by $k=\tan \theta$, which equals to the framing number of the lens space $L(k,1)$. These NS5-branes come from the reduction of D6-branes at two endpoints of the interval $I_6 \in L(k,1)$. Recall that D6-branes are Kaluza-Klein monopoles of the graviton along $S^1_\sharp$, which emerge only at points where the $S^1_\sharp$ shrinks. For lens spaces, only at the endpoints of the interval $I_6$, $S^1_\sharp$ shrinks; see e.g.\cite{Gukov:2015aa} for more details. 
%This property of Lens space naturally engineers NS5-branes. Moreover,  
The D6-brane is T-dual to a D5-brane in IIB string theory. For constructing a gauge theory, what we want is a NS5-brane, so the S-duality in IIB should be applied to turn this D5-brane into a NS5-brane. In short, this process is $\D6 (\IIA)\rightarrow \D5(\IIB) \xrightarrow{{S}} \NS5 (\IIB)$. At the other endpoint of $I_6$, another $\NS5'$-brane emerges analogously, which differs from $\NS5$ by the relative angle $\theta$.

The above gives the 3d brane web $\NS5-\D3-\NS5'$. We need to compare this with various possible 3d $\N=2$ brane configurations
in \cite{Kitao:1999aa} to see if this brane construction is allowed.
In \cite{Kitao:1999aa}, there are three ways to break $\N=4$ to $\N=2$ by turning on various relative angles. Fortunately, the 3d brane web we just obtained is one of them, which is the case that requires the condition $\rho=\theta$, where $\rho$ and $\theta$ are the relative angles of the NS5-branes on planes $x_{9\sharp}$ and $x_{59}$, respectively. Due to this restriction, people often do not distinguish NS5 and D5 and refer to them only as 5-branes, as they differ only by an angle on the plane $x_{59}$. In type IIB string theory, we often set $\theta=\pi/2$, which means that NS5 is along the vertical direction and D5 is along the horizontal direction in the plane $x_{59}$. Hence the $\NS5'$-brane  should be called $(k,1)$ 5-brane\footnote{For a $(p,q)$-brane, $p$ is the electric charge and $q$ is the magnetic charge.}. In addition. these 3d brane webs can also be obtained by Higgsing 5d $\N=1$ brane webs by conifold transitions; see, e.g.\cite{Dimofte:2010tz,Cheng:2021nex,Cheng:2021vtq}.

\subsection{Matter circles and gauge circles}\label{sec:mattercirc}
Matter circles and gauge circles are defined as geometric objects in three-manfolds, corresponding to chiral multiplets and gauge groups, respectively. These two circles can be identified as some circles of the lens spaces, by considering the string dualities and consistency. 
Previously, only the gauge circles are considered, corresponding to the surgical unknots. The matter circle comes from the Oogur-Vafa Lagrangian brane and is the intersection $\redcirc=\L_\redcirc \cap L(k,1)$. In the following, we will discuss how the geometric structure of the lens space determine these two circles.

Basically, since the brane configuration in M-theory is quite simple, one can try all possible choices and finally find only one canonical Lagrangian brane makes sense. This $L_\redcirc$ should take directions $ \R_{34}^2 \times S^1_9 \subset   N_{345} \times L(k,1)_{69\sharp} $ and  leads to a D5-brane, as shown in Table \ref{tab:branedefts}. Other choices may not give rise to meaningful 3d brane webs. Recall that in 3d brane webs, each D5-brane is responsible for a flavor symmetry $U(1)_F$, and the  hypermultiplet is given by the open strings between D3-brane and D5-brane. For $L(0,1)$, what we get is a $\N=4$ theory $U(1)_0$ with a hypermultiplet, while for the lens space $L(k,1)$ with $k \neq 0$, we get $\N=2$ theories.
The evolution of the OV Lagrangian M5-brane under the string duality is $$\M5''(12349) \rightarrow \NS5(12349) (\IIA) \rightarrow \NS5(12349) (\IIB)\xrightarrow{S} \D5(12349) (\IIB) \,.$$
The intersection has to be the longitude $\M5 \cap \M5'' =S^1_9$ of the lens space.
Some discussions on 3d brane webs can be found in e.g. \cite{Kitao:1999aa,Dimofte:2010tz,Cheng:2021vtq}.

There is a subtle problem here. Since T-dualities are only about exchange big circle and small circle,
one may wonder whether the  D5-brane dual to an OV Lagrangian brane should be a line or a circle on the 3d brane webs. 
This problem can be answered by the M-theory/IIB-string duality, using the relation:
\begin{align}\label{radiscomp}
	\frac{l_p^3} {R_{9} R_{\sharp}}= {\tilde{R}_9}\,,
\end{align}
where the radius of the dual circle $\tilde{S}^1_9$ in IIB is infinitely large $\tilde{S}_9^1 \rightarrow \inf$, since the torus area vanishes at both endpoints of the interval $I_6$ of the lens space, namely $R_9R_\sharp \rightarrow 0$. The radius of D5-brane is $\widetilde{R}_9 \rightarrow \inf$ and hence is a line $\R_9$. This means that this M5-brane configuration on the lens space is precisely dual to the standard 3d brane webs, and the dual D5-branes should locate at the two endpoints of the interval which are the locations of the NS5-branes. 

\begin{figure}[h]		
	\centering
				\includegraphics[width=2in]{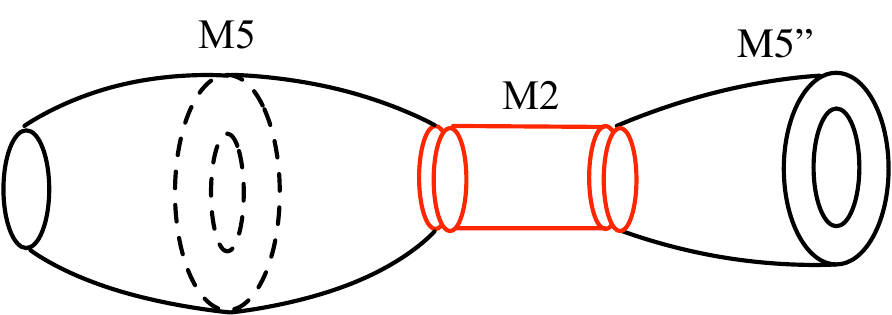}
		\caption{ The  Heegaard splitting of lens space $L(k,1)$ is a torus bundle fibered over an interval, giving by gluing two solid tori. The M2-brane tube has the topology $S^1_9 \times I_5$. }
	\label{fig:defectM52}
\end{figure}

The 3d brane webs could reversely provide more details to the OV construciton. Previouly, we mentioned that the $\L_\redK$ should intersection with the three-sphere $S^3$ and the intersection is just the knot $
\redK$. However, this is not precise. $\L_K$ and $S^3$ should not really intersect, and there is actually a M2-brane stretching between them, and the M2-brane wrap $\redK\times I_5$ is a tube, and  $I_5$ is the distance between $\L_\redK$ and $S^3$.
We illustrate the $\L_\redK$ (wrapped by a $\M5''$) and the three-manifold (wrapped a M5-brane) in Figure \ref{fig:defectM52}, where the flavor symmetry $U(1)_F$ come from the $\M5''$ as it is non-compact, and the gauge symmetry $U(1)_k$ comes from the M5-brane as it is compact. The M2-branes stretching in between give a hypermultiplet in the bi-fundamental representations of the groups $U(1)_k \times U(1)_F$. In this figure, we put the OV M5"-brane on the right endpoint of the interval, which could be moved to the left endpoint through the gluing map of the lens space; this move can also be interpreted in terms of 3d brane web.

From this configuration, we can see that the red circle as the boundary of the M2-brane tube is crucial in describing 3d gauge theories, since it correspond to hypermultiplet, and hence we call it matter circle or BPS states.

The matter circle is the core of the solid torus $S^1_9 \in D^2 \times S_9^1$, and is often referred to as the singular fiber. In the following work \cite{Cheng:2024ybd}, we show that it is useful to emphasize this circle, since it is the key to many 3d dualities. 
The gauge circle  is identified as the meridian $S_\sharp^1$ of the fiber $T^2_{9\sharp}$, which is also the M-theory circle. In the next section \ref{sec:Dehndef}, we will show that this circle is indeed the surgical unknot that determines the lens space. Now, we can see that both the longitude and meridian of the torus are useful.

We should consider how these two circles interact with the gluing map, because the mapping class group can twist and recombine these two circles. Hence a circle on one endpoint will become quite different, when it is moved to the other endpoint, and this movement is governed by the gluing map. 
In \eqref{fig:transport}, we illustrate the transport of both circles from the left endpoint to the right endpoint,  caused by the gluing map in \eqref{gluingmaplen}. It turns out that 
 $S^1_9 =-\tilde{S}^1_9$, and hence the matter circle is unique in the lens space, while the meridian $S^1_\sharp$ is highly twisted.
 \begin{align}
	\centering
	\begin{split}
\includegraphics[width=1.5in]{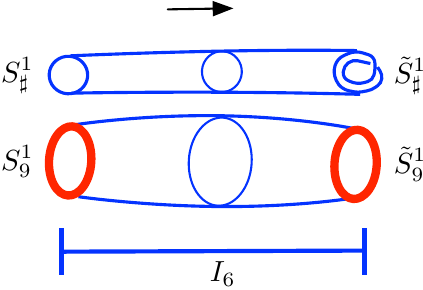}
	\end{split}
	\label{fig:transport}
\end{align}
In terms of brane webs, the gauge circle $S^1_\sharp$ corresponds to NS5-branes, and on the other endpoint we have a $\NS5'$-brane that picks up an angle caused by this twist. D3-brane in between lives the gauge symmetry. The correspondence between a half of the lens space and the 3d brane web is as follows:
\begin{align}\label{hopfobser}
	\centering
	\begin{split}	\includegraphics[width=1.5in]{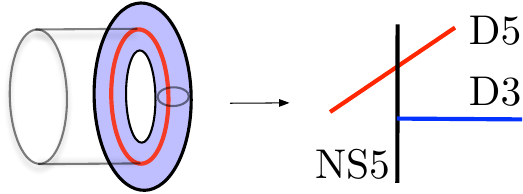}
	\end{split}
\end{align}
where we roughly draw the OV Lagrangian brane as a tube, but remember that its topology is actually a solid torus. This correspondence shows that
gauging the D3-brane is equivalent to gluing a solid tori, or in other words, attaching a NS5 to intersect the D3-brane.

In addition, the $\M5'$-brane in Table \ref{tab:branedefts} is along the meridian $S^1_\sharp$. However, if we perform the S-duality, then $\M5' (1234\sharp)\rightarrow \D5'(12349) \xrightarrow{S} \NS5(12349)$, which unfortunately does not give a D5-brane and hence is not suitable for engineering flavor symmetries. In this case, the M2-brane between the bulk M5 and the defect $\M5'$ is finally dual to a D1-brane: $\M2(5\sharp) \rightarrow \text{F}1(5) \xrightarrow{T_9} \text{F}1(5) \xrightarrow{S} \text{D}1(5)$. The topology of this M2-brane is a tube $S^1_{\sharp} \times I_5$. Even though, this brane is still useful, since it describes an exceptional case for lens spaces with some special framing numbers, as we will shown in \eqref{specialT1}.

\vspace{4mm}
\noindent
\textbf{Mass parameters.}
The M2-brane is dual to the F1-string in 3d brane webs, which is helpful to identify mass parameters. 
The length of the F1 string between D3 and D5 is the real mass parameter, while that of the D1 strings between D3 and NS5 is the FI parameter.  

In term of M-theory,
since the OV Lagrangian $\text{M5}''$ and the compact M5 are separated along the direction $x_5$, the M2-brane should stretch between them, and its one boundary  is the intersection $\M2\cap \M5=S^1_9$, and the other boundary is the intersection $\M2\cap \M5''=S^1_9$. Thus the M2-brane corresponding to the F1 or D1 is a tube $I_5\times S^1_9$ rather than a disk. In short, the M2-brane descends $\M2(59) \rightarrow \D2(59) \xrightarrow{T_{9}} \D1(5) \xrightarrow{S} \text{F}1(5)$ under dualities. 
The mass parameter as the length of $I_5$ takes values in $(-\inf,+\inf)$. If the chiral multiplet is massless, then $x_5=0$. Note that when $S^1_9 \rightarrow 0$, the matter is also massless. We will go back to this parameter in Section \ref{sec:FImassrev} and draw the conclusion from the perspective of topological strings.

\vspace{4mm}
\noindent
\textbf{FI parameters.}
 FI parameter is given by the length of the D1-string stretching between D3 and NS5, which is the relative distance between two NS5 branes. 
Here, we cannot identify the FI parameter by analyzing lens spaces through string duality. We can assume the FI parameter vanishes here, and in this limit lens space and the 3d brane webs can be precisely matched.

If we have to find the FI parameter in the lens space, then it should be $S^1_9$, because it is not shrinkable in the lens space. If one finds a parameter that is not shrinkable in a closed M5-brane, then it should be for a global symmetry, and in this case of 3d theories it should be the topological symmetry, and the associated parameter is the FI parameter.
In Section \ref{sec:FImassrev}, we will use surgery of three-manifolds to verify this parameter is indeed the FI parameter.

\subsection{Plumbed graphs for lens spaces}
In terms of Dehn surgeries, the lens space $L(k,1)$ is represented by a surgical unknot with a framing number, and the corresponding plumbing graph contains one gauge node $\bullet_k$. The above analysis in M-theory shows that the matter circle and the gauge circle form a link as in \eqref{mgcircles}.
This link gives a fundamental chiral multiplet charged under a gauge group
 \begin{align}	\label{lensmatter}
	\centering
	\begin{split}
		\includegraphics[width=1.5in]{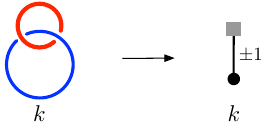}
	\end{split}
\end{align}
where the gray box denotes a chiral multiplet. In the 3d brane webs, a D5-brane leads to a hypermultiplet, and we should decoupling the a half of the hypermultiplet by sending a half of the D5-brane to infinity.

At this stage, the matter circle $\redcirc$ is just the intersection between the OV Lagrangian brane and lens spaces. In Section \ref{sec:STdrill}, we will show that the red circle should be viewed as an identical surgical unknot, where the Rolfsen twist can be applied and leads to $ST$-moves. 

\vspace{4mm}\noindent
\textbf{Ooguri-Vafa configuration revisited.}
Now we have a more precise picture for the Ooguri-Vafa construction, inspired by 3d brane webs.
	The OV Lagrangian brane is $\L_\redcirc=\redcirc \times \R^2$ where $\R^2$ is a section in the fiber of $T^*M_3$. The intersection $\redcirc = \L_\redcirc \cap M_3$ is the matter circle. Topologically, $\L_\redcirc$ can be written as a solid torus $\redcirc \times D^2 = T^2 \times \R_+$ where $T^2$ degenerates to the circle $\redcirc$ at the origin of $D^2$. We illustrate the OV construction below
\begin{equation}\label{fig:OVdefect}
	\begin{aligned}	\includegraphics[width=1.2in]{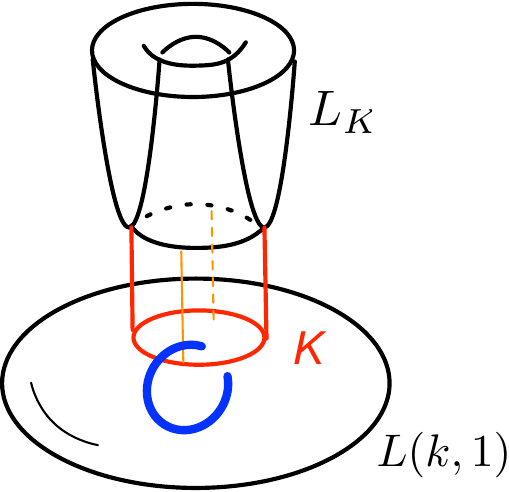}
	\end{aligned}
\end{equation}
Note that $\L_{\redK}$ does not really intersect the lens space, as red tube $\redK \times I$ wrapped the M2-brane separate them, and only when $I \rightarrow 0$, they could intersect along the knot $\redK$. At this limit, the chiral multiplet is massless.
The plumbing graph \eqref{fig:OVdefect} just represents the Figure \ref{fig:defectM52}.

\vspace{4mm}\noindent
\textbf{Many chiral multiplets.}
If many $\L_{\redcirc}$ is inserted in the bundle, they will share the same circle $S^1_9=\redcirc$ and are distributed as parallel planes on the direction $x_5$ in the fiber of the bundle. We can distinguish them by giving different mass parameters $m_i$ (the coordinates of the OV M5-branes on the direction $x_5$) and different charges as winding numbers $q_i$ around $S^1_9$. Lens spaces with OV M5-branes can engineer a class of theories\footnote{Note that $\AF$ differs from $\F$ by the opposite charge.}:
\begin{equation}
	U(1)_k+ \sum_i N_{f,q_i} \F_{q_i}+  \sum_i N_{f,q_i}\AF_{-q_i} \,.
\end{equation} 
When these OV M5-branes coincide, flavor symmetries enhance to $U(N_f)$, corresponding to the overlapped D5-branes.

\section{Brane webs}\label{sec:branewebs}
The M5-brane wrapping the lens space $L(k,1)$ is dual to the 3d brane web $\NS5-\D3-(k,1)\text{5-brane}$.
In this section, we discuss various properties of the 3d brane webs, which surprisingly could be translated as the geometric properties of lens spaces, and hence deep the understanding on both sides.

\subsection{D5-brane move}

\begin{figure}[h]
	\centering
\includegraphics[width=3.5in]{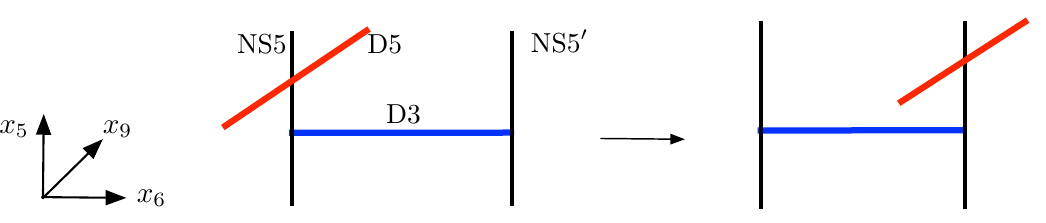}
\caption{Move D5-brane from one endpoint to another endpoint of the D3-brane.}
	\label{fig:moveD5}
\end{figure}

In 3d brane webs, D5-branes can be moved from one NS5-brane to another NS5-brane as shown in Figure \ref{fig:moveD5}.
This movement do not change the 3d gauge theories, and can be interpreted using to the Heegaard splitting of the lens spaces \eqref{fig:transport}. What will happen to the D5-brane during this move is determined by the gluing map  \eqref{gluingmaplen}. Explicitly, $S_9^1$ and $S^1_{\sharp}$ are mapped to $\tilde{S}_9^1$ and $\tilde{S}^1_{\sharp}$ respectively,
where
$\tilde{S}_9^1= -S^1_9 $ and $\tilde{S}_\sharp^1=k S^1_9 +  S^1_\sharp$. 
 The $\D5$-brane is almost invariant under this movement, since $\tilde{S}_9^1=-S_9^1$  only flips the sign of the charge by reversing the orientation. This move is also discussed in \cite{Cheng:2021vtq}.

\subsection{An exceptional brane webs}\label{sec:exceptionalcase}
The lens spaces $L(0,1) =S^2 \times S^1$ engineer abelian 3d $\N=4$ theories, and the corresponding 3d brane webs have two parallel NS5-branes. 
In the presence of the D5-brane, the hypermultiplet is introduced to give the theory $U(1)_k+1 \F +\AF$.
We should decouple the antichiral matter $\textbf{AF}$ by sending its mass $m_a \rightarrow -\inf$, using the formula $k_{\text{eff}} = k +\big(N_f \text{sign}(m_f)   - N_a \text{sign}(m_a) \big)/2$ for the theory $U(1)_0+N_f\textbf{F}+N_a\textbf{AF}$, and eventually this decoupling leads to $k_{\text{eff}} = \pm 1$ and the theory becomes $
U(1)_{\pm1}+ 1\textbf{F}$. This is the basic theory shown in $ST$-moves. 

Let us look at the brane webs from the direction $x_6$ for convenience. The decoupling process is equivalent to splitting a D5-brane into two parts and the part for the anti-chiral multiplets is sent to infinity and hence disappear.
\begin{equation}\label{webplus1}
	\begin{aligned} \includegraphics[width=2.5in]{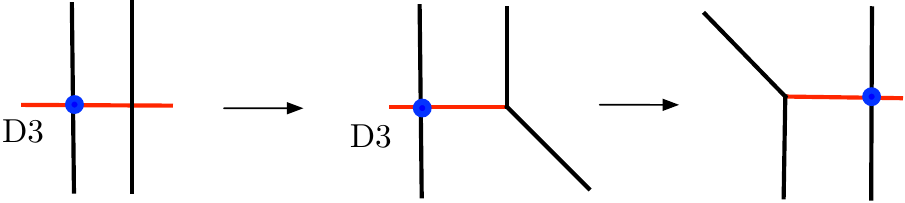}
		\end{aligned}
	\end{equation}
	where blue nodes denote D3-branes along the direction $x_6$, red lines denote D5-branes, and vertical lines denote NS5-branes.
The brane webs for $U(1)_{\pm1}$ with a D5 brane can reduce to the theory $\bullet_{\pm1}-\graybox$ as follows: \begin{equation}\label{L11webs}
	\begin{aligned} \includegraphics[width=2.5in]{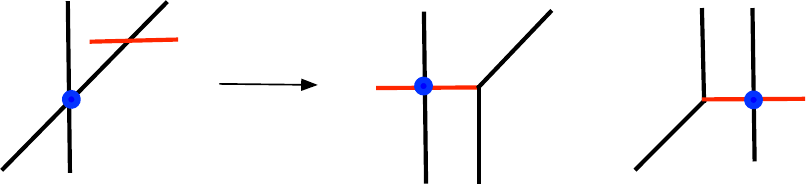}
	\end{aligned}
\end{equation}
where a half of the D5-brane leads to a massless chiral multiplet.

There is a special case that the brane web for $U(1)_{\pm1}$ with an additional NS5-brane can also lead to $\bullet_{\pm1}-\graybox$ as follows:
\begin{equation}\label{specialT1}
	\begin{aligned} \includegraphics[width=4in]{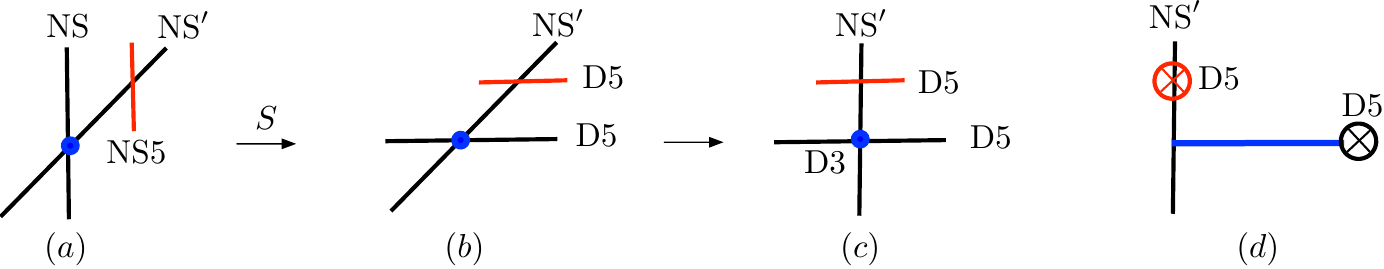}
	\end{aligned}
\end{equation}
where from $(a)$ to $(b)$, we apply the S-duality, and from $(b)$ to $(c)$ we just change the slope of $(b)$ by the transformation $SL(2,\mathbb{Z})$, which turns the $(p, 1)$-brane to the $(p+n, 1)$-brane, but preserves the $(1,0)$-brane. $(d)$ is the same web as $(c)$ and we just see the brane web from a different direction.

 Obviously, $(c)$ is equivalent to the first brane web in \eqref{webplus1} by the S-duality. Because of this, the NS5-brane corresponding to an Ooguri-Vafa M5-brane on $S^1_\sharp$ can also lead to $\bullet_{\pm1}-\graybox$ after decoupling. This is an exceptional case, since the NS5-brane comes from $S^1_\sharp$ rather than $S^1_9$ that we used before.

 Similarly, the 3d brane webs can be drawn for $U(1)_{1/n} + 1\NS5$ for any $n \in \mathbb{Z}$, and they all end up with the same web $(d)$ up to $SL(2, \mathbb{Z})$ transformations. Thus, $U(1)_{1/n} + 1 \NS$5 are also admitted and are equivalent for any integer $n$. 
This phenomenon looks a surprise for 3d brane webs and 3d gauge theories, but can be naturally understood through properties of three-manifolds. The lens space $L(1,1)$ have have equivalence $L(1,n)=S^3$ for $\forall n \in \Z$, and different $n$ are related through equivalent Dehn surgeries, and we will give more details in section \ref{imagecircs}.

\subsection{Brane webs on torus}\label{sec:branewebtorus}
In this subsection, we revisit another construction of 3d brane webs in \cite{Kitao:1999aa}, where Lagrangian M5-branes wrapping some circles in the torus $T^2_{9\sharp}$ are dual to 5-brane webs, and M2-branes between M5-branes lead to the D3-brane, which is quite different from the construction that we discussed above. The benefit of this construction is to give brane webs with generically charged chiral multiplets. This subsection can be skipped if one is not interested in brane webs.

Basically, M5-branes do not wrap the whole torus, but intersect the torus along some circles $qS^1_9+pS^1_\sharp$, and the M2-branes stretching between two M5-brane webs are horizontal lines $\{\star\} \times I_6$ in the domain $ T^2_{9\sharp}\times I_6$.
On the two boundaries of the domain $T_{9\sharp}^2 \times I_6$, we should generate two 5-brane webs realized by the Lagrangian M5-branes  wrapping the $(q,p)$ torus knots on $T_{9_\A\sharp}^2$, which leads to the $(q,p)5$-brane in IIB string theory.
 In addition, M2-branes wrappings $q$ times along $S_9^1$ and $p$ times along $S_\sharp^1$ leads to the $(p,q)$ string in type IIB string theory. However, the M2-brane stretching between two M5-branes  is dual to the D3-brane. In this way, this construction gives the standard 3d brane webs and more.

\begin{figure}[h]
 	\centering
 	%   \begin{split}
 	\includegraphics[width=3.5in]{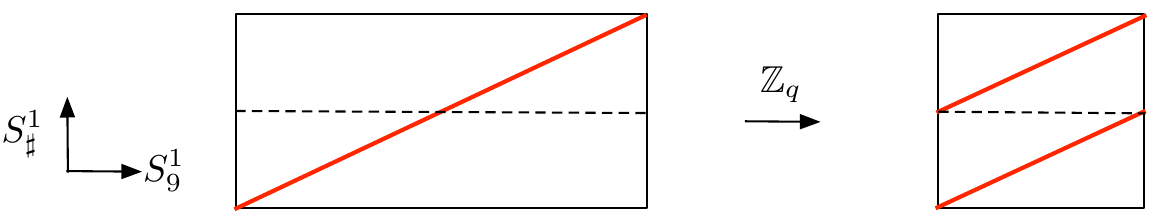}
 	%    \end{split}
 	\caption{In this example we choose $q=2$. The charge $q$ equals to the number of components distributed along meridian $S^1_{\sharp}$. The red line is the circle wrapped by the M5-brane. The quotient is given by $S_\sharp^1 \rightarrow S^1_\sharp\,,~S_9^1 \rightarrow q S_9^1$.}
 	\label{fig:qtwomatterbrane}
 \end{figure}

Let us discuss how to realize the charge $q$ hypermultiplet. We coincide $q$ copies of D5-branes to get a $(q,0)$-brane, and correspondingly in M-theory we wrap a M5-brane $q$ times along the circle $S^1_9$ to realize it, as illustrated by red lines on the torus:
\begin{align}\label{manyredlines}
	\centering
	   \begin{split}
	\includegraphics[width=0.8in]{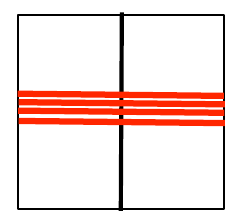}
	   \end{split}
\end{align}
Then if a D3-brane ends on the charge-$q$ D5-brane, it has $q$ choices as Higgs vacua, since the  M5-brane wrapping the circle $q S^1_9$ has $q$ separate lines. These $q$ vacua are points evenly distributed along $S^1_\sharp$, which separate the M5-branes into $q$ fractions as shown in \eqref{manyredlines} and Figure \ref{fig:qtwomatterbrane}. The relative real mass parameters between the fractions are $m_i \sim S^1_\sharp/q$. 
 One can take a quotient $\Z_q$ on the torus to understand this winding number $q$, as shown in Figure \ref{fig:qtwomatterbrane}. 
 
If we turn  this domain $T^{2}_{9\sharp}\times I_6$ into a lens space by letting the $S^1_\sharp$ shrinks at endpoints of the interval $I_6$, these fractional lines coincide and become an overlapped charge-$q$ D5-branes. There is still an overall mass parameter $m_0$ left, which is the length of the F1(5) string stretching between the D3-brane and the charge-$q$ D5-brane.

\vspace{4mm}\noindent
\textbf{Decoupling.}
If a NS5 meets a charge $q$ D5, one can split this D5 into two half-infinite long D5-branes, and each conponent leads to a chiral multiplet of charge $q$ or $-q$, depending on their relative positions with D3-branes. The effective CS level is given by the relative angle between two  NS5-branes, which can be quantum corrected by D5-branes:
\begin{equation}\label{qutshift}
\Delta k = \tan \theta = \frac{q^2}{2}\text{sign}(m_{\F_{q}}) - \frac{q^2}{2}\text{sign}(m_{\AF_{-q}}) \,.
\end{equation}
 When a NS5 meets a charge $q$ D5-brane component, the brane could merged and pick up an additional charge $(q,0)$ due to charge conservation. In total, there are $q$ D5-brane fractions, and then the $(0,1)$-brane is finally turned into a $(q^2,1)$-brane, as shown in the following:
\begin{align}\label{multiplewebs}
	\centering
	\begin{split}
		\includegraphics[width=1.7in]{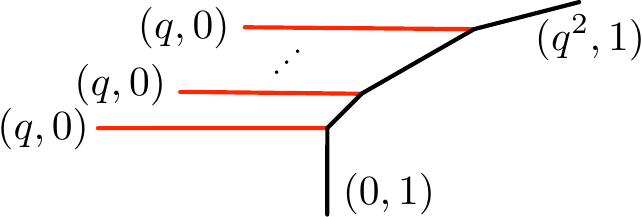}
	\end{split}
\end{align}
The other half-infinite 5-branes pointing to the right and lead to corrections as well. We do not repeat the description here.
The shift $\Delta k$ depends on the signs of mass multiplets. In particular, if $\Delta k= \pm q^2$, then $ \text{sign}(m_{\F_{q}}) \cdot \text{sign}(m_{\AF_{-q}})  =-1 $, and if $\Delta k= 0 $, then $ \text{sign}(m_{\F_{q}}) \cdot \text{sign}(m_{\AF_{-q}})  =+1 $.  See \cite{Cheng:2021vtq} for this subtle detail.
The brane web in \eqref{multiplewebs} has not been proposed before but is useful, since it produces the right formula for the effective Chern-Simons levels in \eqref{qutshift}. 

We are interested in the theory with charge $q=+1$ chiral matter, which is just the theory $\bullet_{+1}-\graybox$.  Decoupling $\AF$ changes the slope of NS5 as follows:
 \begin{align}
	\centering
	\begin{split}
		\includegraphics[width=1.5in]{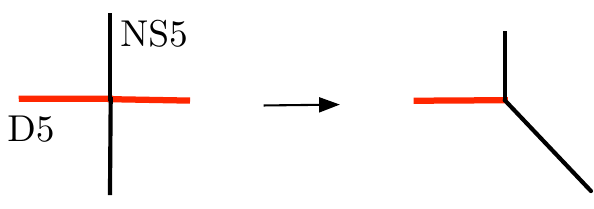}
	\end{split}
\end{align}
After decoupling, the circle $S^1_\sharp$ fuses with $S^1_9$ to give $S_\sharp^1 \rightarrow S_{\sharp}^1\pm S_9^1$. This transforms $L(0,1)$ to $L(\pm 1,1)$, which are not equivalent, and hence the decoupling changes the three-manifold significantly.

To some extent, we can use the lens spaces to understand $S_\sharp^1 \rightarrow S_{\sharp}^1\pm S_9^1$ that causes the geometric change from $L(0,1)$ to $L(\pm 1,1)$. Since the 3d brane web corresponds to a lens space with an OV Lagrangian M5-brane, this geometric change is
\begin{align}\label{decoupAFtoS3}
	( S^1 \times S^2 ) \sqcup L_\redcirc ~\xrightarrow{\text{decoupling \AF}} ~ S^3 \sqcup L_\redcirc  \,.
\end{align}
Moreover, we can only decouple $\AF$ by taking $m_a\rightarrow -\inf$ and keep $\F$, but this does not change $k_\eff=\pm1$ since $\F$ is assumed to have a positive mass in the renormalization of 3d theories; see \cite{Tonglec} for a nice discussion. This geometric change caused by decoupling is analogous to a particular Dehn twist, but not the equivalent one. 
After this inequivalent twist, the framing numbers describing effective CS levels change by $\bullet_k \rightarrow \bullet_{k+q^2}$. We will go back to this topological change in the paragraph below \eqref{S2S1tw}.

\vspace{4mm}\noindent
\textbf{Another brane configuration.}
This brane configuration used in \cite{Gadde:2013wq}  is similar to the one we have just discussed, but there is a little difference but  is not essential.
In this configuration, Lagrangian M5-branes along $(012349)$ and $(01235\sharp)$ are introduced in the cotangent bundles of the domain $T_{9\sharp}^2\times I_6$. These Lagrangian M5-branes could combine into a single M5-brane in M-theoy.
To be more precise, the Lagrangian M5-brane wraps $p S^1_{\sharp}+q S^1_{9}$ is dual to a $(q,p)$ 5-brane in IIB string theory.
The 5d brane web is on the plane $(45)$, and the spacetime of the 5d theory spacetime is $(01239)$. This configuration is more natural to describe the 5-brane webs for 5d gauge theories. 
The combined M5-brane and the dual 5-brane webs is illustrated below
 \begin{align}\label{torusweb}
	\centering
	\begin{split}
	\includegraphics[width=2.5in]{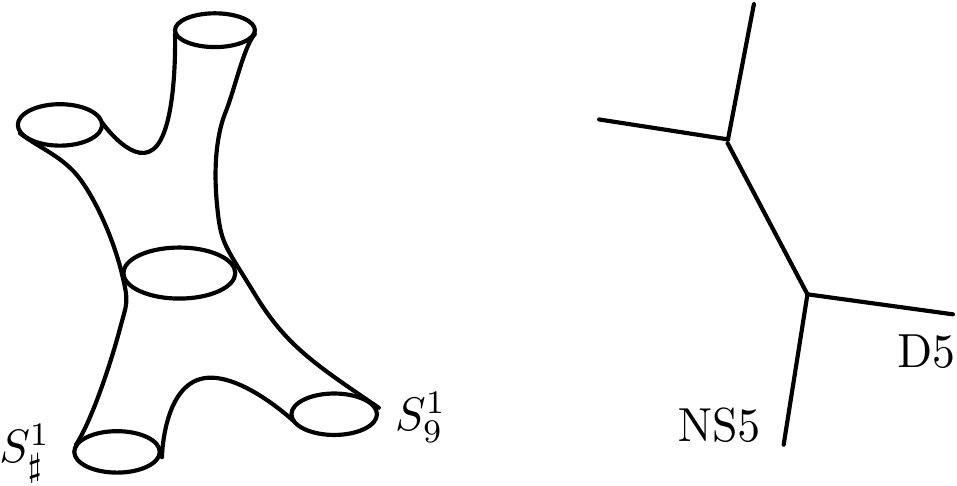}
	\end{split}
\end{align}

To generate 3d brane webs, we have two choices. We can add
a M2-brane along $(036)$, which is dual to a D3-brane along $(0369)$, and  the 3d theory lives on $(039)$. 
Another choice is that  we can wrap a M5-brane on the three-manifold $T_{9\sharp}^2 \times I_6$, which also leads to a D3-brane along $(0126)$ and the 3d theory lives on $(012)$.

\section{Plumbed three-manifolds }\label{sec:matterinthree}

Abelian 3d gauge theories labeled by plumbed three-manifolds are considered in \cite{Gadde:2013aa}.  However, chiral multiplets have not been considered for these three-manifolds. 
Lens spaces and Lagrangian M5-branes that we have discussed could more or less fill this gap. One may think about using 3d brane webs to describe the plumbed three-manifolds, but it is not practical, since only very few 3d brane webs are known. We can rely on surgeries to analyze the Lagrangian branes and Kirby moves. In this section, we review surgeries and then discuss their properties.

\subsection{Dehn surgeries}\label{sec:Dehndef}
Plumbed three-manifolds are defined by Dehn surgeries over the link $L_1 \cup L_2 \cup \cdots\cup  L_n$ in the three-sphere $S^3$. Plumbing graphs is a short hand notation for the surgical links by $L_i \rightarrow \bullet_{k_i}$ and $L_i \cup L_j \rightarrow  \bullet_{k_i}- \bullet_{k_j}$. 
Surgery is powerful, since any closed-oriented three-manifold $M_3$ can be obtained by
drilling out the neighborhoods of links and then filling in solid tori:
\begin{equation}
M_3 =S^3(L_1 \cup L_2 \cup \cdots\cup  L_n) :=
\left(  S^3- \cup_{i=1}^n N(L_i)  \right) ~\sqcup_{f_i}~  \left( \oplus_{i=1}^n D_i^2 \times S_i^1 \right)\,,
\end{equation}
which is illustrated in Figure \ref{fig:surgeryKde}. 
%$\partial(S^3- N(L_i))= T^2$. 
The gluing map ${f}_i$ maps the boundary of the $i$-th solid torus to the boundary of the $i$-th link complement, which introduces a framing number $f_i =p_i/q_i$, as we saw in the case of lens spaces.  Note that these solid tori $D_i \times S^1$ are independent. In addition, $S^3-\cup_i N(L_i) $ is called the knot complement and is often denoted by $S^3\backslash L_1 \cup L_2 \cup \cdots  \cup L_n$. 

The gluing map $f_i$ maps the meridian $\a_i$ of the torus boundary of $D^2_i \times S_i^1$ to a curve $J_i$ on the boundary of the $i$-th complement
% which is the boundary of circle complement. 
by $f_i(\a_i)=J_i$, which is sufficient to determine the surgery on the unknot $L_i$. The meridian $\a_i$ and the Jordan curve $J_i$ play the role of gauge circles, since they correspond to the shrinking circles and hence NS5-branes for lens spaces.
The images of the meridian and longitude for the $i$-th solid torus are given by
\begin{align}\label{gluingmaplenDehn}
	\begin{bmatrix}
		f_i(\a_i)  \\
		f_i(\b_i)
	\end{bmatrix}=
	\begin{bmatrix}
		k_i & -1\\
		1 &0
	\end{bmatrix} \cdot
	\begin{bmatrix}
		{\a}_i'  \\
		{\b}_i' 
	\end{bmatrix}
	\,,
\end{align}
where $\a_i'$ and $\b_i'$ denote the meridian and longitude on the boundary of the $i$-th unknot complement, and the
 Jordan curve is $J_i= k_i{\a}_i' -{\b}_i'$. The framing number is given by $L_i\cdot L_i := \b_i' \cdot J_i=k_i $, so the longitude ${\b}_i'$ could represent the surgical unknot $L_i$. In the following, we show the locations of these circles on the complement $S^3 \setminus L_i$:
 \begin{align}\label{abprime}
	\centering
	\begin{split}
		\includegraphics[width=1.2in]{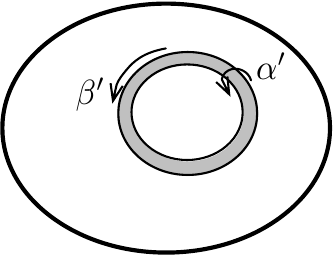}
	\end{split}
\end{align}

 \vspace{4mm}\noindent
\textbf{Bifundamental chiral multiplet.}
The lens space is given by the surgery along a unknot $L(k,1) = S^3_{k}(\bigcirc)$ with a framing number $k$. When the 
Ooguri-Vafa Lagrangian brane is present, we get a  fundamental chiral multiplet. Kirby moves could act on the surgical unknot and blow it up to a complicated plumbing graphs.

To get chiral multiplet in other representations, we should link the OV Lagrangian branes to a number of surgical circles. For instance a
bifundamental chiral multiplets is given by linking a matter circle  to two gauge (surgical) circles.
For these chiral multiplets, the charge $q_i$ is determined by the linking number between the matter circle and the $i$-th gauge circle. 
An interesting point is that through the $ST$-move, the bifundamental chiral multiplet becomes a fundamental chiral multiplet, and the price to pay is that the mixed Chern-Simons level is changed, as shown in \eqref{bibeta}. 
In addition, the fundamental matter becomes massless after $ST$-moves \cite{Cheng:2023ocj}.
We should an example for the bifundamental chiral multiplet below, whose $ST$-move can be obtained analogously to that of the fundamental chiral multiplet,
\begin{align}\label{directbicirc}
	\centering
	\begin{split}
		\includegraphics[width=2in]{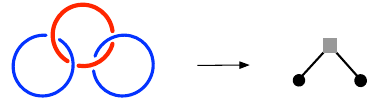}
	\end{split}
\end{align}

%===================

\subsection{Equivalent surgeries and matter circles}\label{imagecircs}
The surgery of each unknot $L_i$ is determined by only the image of the meridian $\a_i \mapsto J_i$.
However, this nice property does not apply to the matter circle, since the matter circle is along the singular fiber (the core) of each solid torus, which is more like the longitude of the boundary of the solid torus rather than meridian. When the solid torus attached with an OV Lagrangian brane is filled to a boundary of the surgical link; see Figure \ref{fig:surgeryKde}, the image of the matter circle becomes a subtle problem. Does it link to the surgical circle or just parallel to the surgical unknot? The answer can be provided by analyzing gluing maps.

In the previous sections, we mentioned that the gluing maps should correspond to the move of D5-branes between two NS5-branes in 3d brane webs. Indeed, there is a nice dictionary between this move and the gluing map. Various properties of 3d brane webs shown in Section \ref{sec:exceptionalcase} can be interpreted by gluing maps. 

To answer these questions and see conclusions we should analyze gluing maps for lens spaces, which is already sufficient, since  we can extend the results to plumbed three-manifolds. This section is parallel to Section \ref{sec:exceptionalcase}. In the following, we give the technical discussion.

\vspace{4mm}
\noindent
\textbf{Lens spaces ${L(k,1)}$.}
The lens space is given by $ L(k,1):=(S^3\setminus \bigcirc) \cup_{f_k} (D^2 \times S^1) $. 
We denote the meridian and longitude on the boundary of the solid torus as $(\a,\b)$, and that on the boundary of the complement $S^3\setminus \bigcirc$ are $(\a',\b')$.
We denote their images as $\Ima=f(\a)$ and $\text{Im}(\b)=f(\b)$, and then the gluing map is represented by a matrix in $SL(2,\Z)$. 
\begin{align}\label{glueformula}
	\begin{bmatrix}
		f(\a) \\
	\fb
	\end{bmatrix} =
	\begin{bmatrix}
		p & -q \\
		r & s 
	\end{bmatrix} \cdot
	\begin{bmatrix}
		\a' \\
		\b'
	\end{bmatrix} \,,
\end{align}
where $ ps+qr=1$, and the Jordan curve is $J = \text{Im}(\a) = p\a'-q\b'$.

For the lens space $L(k,1)$, we have $p=k$, $q=1$, and $r=-sk +1$ for any $s \in \mathbb{Z}$.
The images of gauge circle and matter circle are given by
\begin{equation}\label{lenskimage}
\begin{aligned}
	&\fa =  k \a'  -\b' \,,\\ 
	&\fb= (-sk+1) \a'+s\b' =  \a' - s( k \a' -\b')  = \a' -s \fa  \,.
\end{aligned}
\end{equation}
Notice that the image $f(\b)=\text{Im}(\b)$ contains $s$ copies of $f(\a)$. We can fix the free integer $s$  by setting $s=0$ and then $f(\b)=\a'$. This means that the matter circle winds the meridian of the complement boundary and hence link to the surgical circle $\b'$. 
 In short, $(\fa,\fb) = ( k \a'-\b', \a')$. This just describes the link graph in \eqref{lensmatter}, where the surgical circle link to the matter circle.
 
The special case is when $s=\pm1$ and $k=\pm1$,  the image of the matter circle is $\fb=\pm\b'$, which means that the matter circle is  the longitude of the surgical unknot. Correspondingly,
this means that in the 3d brane web a D5-brane at one endpoint of the D3-brane transforms to a NS5-brane at the other endpoint, as the surgical unknot is shrinkable in a lens space. This is the exceptional 3d brane web shown in \eqref{specialT1}.

Now, we can see that through adjusting the free parameter $s$, the image of the matter circle  in $L(\pm1, 1)$ can be either $\a'$ or $\b'$.
The integer $s$ can also be fixed in another way. Note that the linking number of image circles is $\fa \cdot \fb =2 s k-1$.  One can set $s=0$, such that $\fa \cdot \fb=-1$. For the special case, we have $\fa \cdot \fb =1$ by setting $s=\pm1$. Both cases  ensure that the meridian has an intersection number $\pm1$ with longitude, avoiding the definition problem.

\vspace{4mm}
\noindent
\textbf{Three-sphere $\mathbf{S^3_\inf}$.}\label{sec:threespheren}
This is the three-sphere with an infinite framing number and is denoted by
$L(1,0)=S^3_\inf$. Its gluing map \eqref{glueformula} is
\begin{equation}\label{spheremap}
	(\fa, \fb) = (\a', r\a'+\b' ) \,.
\end{equation}
 $L(1,0)$ defines the identical surgery along the circle $\bigcirc_\inf$, obtained by filling in the same solid torus that was drilled out, so $(\a,\b) \mapsto (\a',\b')$ and the free integer $r=0$ is fixed.

We think of $S^3_\inf$ as gluing two solid tori in M-theory, and the dual 3d brane web in type IIB string theory exhibits a non-trivial phenomenon under the move of 5-branes: $\NS5 \rightarrow \D5$ and $\D5 \rightarrow \NS5$, namely the NS5 and D5 are switched on endpoints. We use the brane web below to illustrate:
\begin{equation}\label{S3infweb}
	\begin{aligned}
		\includegraphics[width=1.7in]{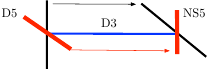}
	\end{aligned}
\end{equation}
Therefore, identical surgery also leads to the exceptional brane web. The theory given by \eqref{S3infweb} is the self-dual 3d $\N=4$ $U(1)$ theory with a hypermultiplet, as shown in \eqref{specialT1}. We will show in the next section that the identical surgery and the exception case $L(\pm1,1)$ are equivalent up to a Rolfsen twist.
 
 \vspace{4mm}
\noindent
\textbf{Three-sphere $\textbf{L(1,n)}$.}
Recall that $L(1,n)$ for any $n\in \mathbb{Z}$ is the same three-sphere $S^3$, providing infinitely many equivalent framing numbers. A special case of $L(1,n)$ is $L(1,0)$. Let us show that the framing number $1/n$ leads to a special property, and this $n$ is the number of Dehn twists applied on the boundary of the complement $S^3\setminus\bigcirc$ by $\a'= \a'' + n\b''\,, \b' = \b''$. This provides equivalent surgeries,  called Rolfsen twists \cite{Rolfesn:1984}.
Taking this twist into \eqref{spheremap}, we get the images:
\begin{equation}
\begin{aligned}\label{sphereimage}
	%\text{image of matter circle:} ~~
	\fa &= \a'' + n\b''\,,   \\
	\fb &= r \a'' +(r n +1) \b'' =r (\a''+n\b'') + \b''   = r \fa +\b'' \,,
\end{aligned}\end{equation}
 Once again, $f(\b)$ contains $r$ copies of $f(\a)$ and one can choose $r=0$ to make it simple. Then we have $\fb=\b''$ for any $n$, and the image of the matter circle is the longitude of  complement $S^3\setminus \bigcirc$ and hence is parallel to the surgical circle $\bigcirc$.
This twist can be reversed and allow us to return the $L(1,0)$ by 
applying the Dehn twist $T_{\a''}$: $\b' = \b''+t \a''\,, \a'=\a'' $, which cancels the integer $r$ in \eqref{spheremap} if setting $t=-r$. More details about equivalent surgeries can be found in textbooks on three-manifolds.

The equivalent $L(1,n)$ for different $n$ can be interpreted through the $SL(2,\mathbb{Z})$ transformation and S-duality of 3d brane webs. 
The brane web for $L(1,n)$ can be translated into the special brane webs shown in subsection \ref{sec:exceptionalcase}, 
where the special 3d brane webs  show the equivalence for different relative angles $\tan \theta =1/n$.

An observation is that \eqref{lenskimage} and \eqref{sphereimage} are equivalent if the meridian and longitude are exchanged. However, $L(k,1)$ and $L(1,n)$ are absolutely different and there is no exchange symmetry between them. We should always distinguish the meridian and longitude, and do not randomly switched these two circles.

\vspace{4mm}
\noindent
\textbf{$\mathbf{S^2 \times S^1}$.}
This geometry is given by $L(0,1)$, and is often used for the connected sum of two three-manifold. If using $\bullet_0$ to denote this manifold, the connected sum of two lens spaces is  $\bullet_{k_1} -\bullet_0-\bullet_{k_2} = \bullet_{k_1+k_2}$.
The images of meridian and longitude are
\begin{equation} \label{S2S1tw}
	(\fa, \fb) = (-\b', \a'+s\b' ) \,.
\end{equation}
This manifold corresponds to the 3d brane web in Figure \ref{fig:moveD5}.
We should set $s=0$ such that the D5-brane does not change when moving it from one NS5-brane to the other NS5-brane. If not fixing $s$, we can 
perform the Rolfsen twist $ \a'=\a''+t\b''$ and $\b'=\b''$, so that the term $s \b'$ can be canceled by setting $t+s=0$. 

However, if we perform the Dehn twist $ \a'=\a''$ and $\b'=\b''+m\a''$, then $ \fa = m \a''-\b''\,, \fb= (-s m+1)\a''+s \b''$. This turns $L(0,1)$ into a lens space $L(m, 1)$, so this Dehn twist is not an equivalent twist, but still is very useful, since it describes the decoupling of chiral multiplets as shown in \eqref{decoupAFtoS3}, where the geometric change happens due to an inequivalent  Dehn twist $\b'=\b''+m\a'' $, which is just the geometric change given by decoupling $S^1_\sharp \rightarrow S^1_\sharp + m S^1_9$.  
This example indicates that both equivalent and inequivalent Dehn twists play useful roles.

\vspace{4mm}
\noindent
\textbf{$\mathbf{L ( \pm1,1)}$.} As we have shown before, $L(\pm 1,1)$ is a very special case that the image of the matter circle $\b$ can be on either meridian or longitude of the complement, and these two locations are equivalent.
This motivates us to guess that the Rolfsen twist should be interpreted as the $ST$-move that exchanges these two locations. We will discuss this in section \ref{sec:STdrill}.

\vspace{4mm}
In the above, we discuss the images of the circles in the complements of various types, and show the correspondence with the special 3d brane webs through Dehn twists.
Let us summarize the images of circles of various types of lens spaces in Table \ref{Tab:imagecirc}\footnote{If one translates surgery of lens space as gluing two solid tori, then the torus switch $\a' \leftrightarrow \b'$ should be applied.}.
\begin{table}[h!]	
	\centering
	\text{
		\begin{tabular}{|c||c| c | c | c  | c |}
			\hline 
			solid torus $D^2\times S^1$	
		& $L(k,1)$ & $L(1,n)$  & $L(\pm1 ,1)$ & $L(1,0)$ & $L(0,1)$  \\ \hline
$\Ima$ of meridian:  $\a$ & $k \a'-\b'$ & $\a'+n\b'$ & $\a'\pm\b'$ &$\a'$  & $-\b'$ \\ \hline 
$\Imb$ of longitude: $\b$  & $\a'$&  $\b'$ & $\a' \xleftrightarrow{\text{twist}} \b'$ & $\b'$ & $\a'$ \\ \hline 
	\end{tabular}		}
	\caption{ 
		The locations of gauge circles and matter circles for various lens spaces.
		}
		\label{Tab:imagecirc}
\end{table}

\subsection{Cobordisms}\label{sec:cobordiam}
 In this section, we represent three-manifolds as cobordisms, which are helpful to understand the locations of M2-branes . 
The red circle $\redcirc$ is the core of the solid tori should be inserted into three-manifolds. It is hard to imagine how M2-branes could end on  circles insider three-manifolds. We notice that cobordisms provide a good picture by representing three-manifolds as fiber bundles over one-dimensional trees. For example,
the solid torus $D^2 \times \redcirc$ wrapped by a Lagrangian M5-brane can be denoted by:
\begin{equation}\label{solidtorus}
	\begin{aligned}
	\includegraphics[width=1in]{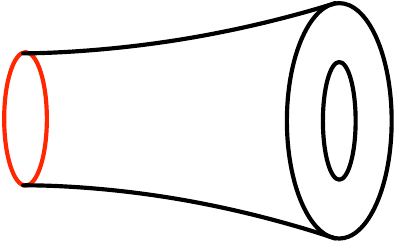}
	\end{aligned}
\end{equation}
which is a torus bundle over an interval, and at the endpoint (the origin of the disc $D^2$) $S^1_\sharp$ shrinks but $S^1_9$ does not. 
Similarly, a three-manifold locally looks like a cobordism. Adding an OV Lagrangian M5'-brane that links to a surgical unknot of a lens space is illustrated by the following graph:
\begin{equation}\label{3mfdcobordism}
	\begin{aligned}
		\includegraphics[width=2.5in]{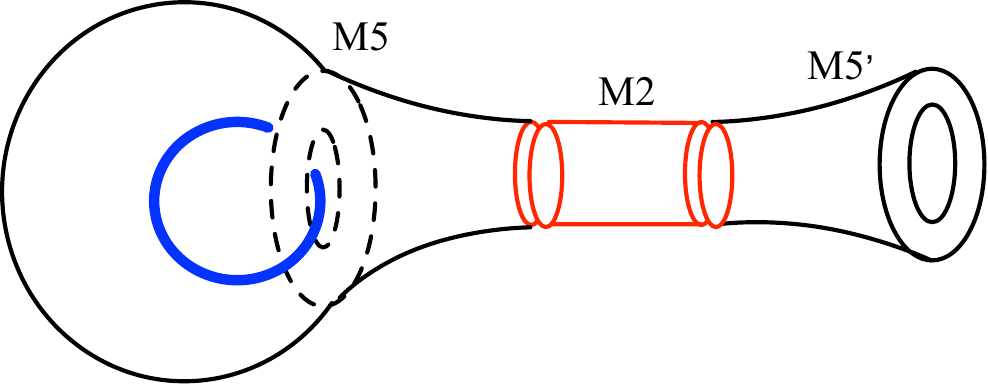}
	\end{aligned}
\end{equation}
where the blue circle denotes the surgical circle. This graph is just \eqref{fig:OVdefect}. 
In generic plumbing graph, each component of the surgical link can be stretched and locally looks like a cobordism, and the surgical unknot (blue) is linked to the red circle. 

The graph in \eqref{3mfdcobordism}  is not a torus decomposition of three-manifolds. The topology of the M2-brane here is $S^1 \times I$ rather than $T^2 \times I$. It would be interesting to find the relations between cobordisms and torus decompositions of three-manifolds.

\vspace{4mm}\noindent
\textbf{Hopf link.}
Another example is the Hopf link. To obtain it, we drill out two solid tori to get the Hopf link complement. After drilling the first solid torus along $L_1$, we get a complement which is equal to a solid torus if its boundary was taken to the outside, namely $S^3\backslash L_1=D^2\times S^1$. In this process, the meridian and longitude are exchanged, which is called  torus switch. Drilling the neighborhood of $L_2$ leads to a torus hole inside, as illustrated below
\begin{equation}\label{hopftorus}
	\begin{aligned}
		\includegraphics[width=2in]{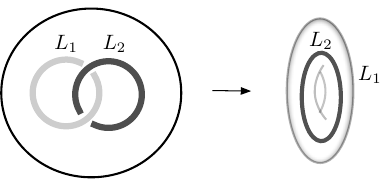}
	\end{aligned}
\end{equation}
Using this figure, one can identify the relations between meridians and  longitudes of $L_1$ and $L_2$.
Its cobordism is as a foliated bundle and each leaf is a torus:
\begin{equation}\label{cobor}
	\begin{aligned}
		\includegraphics[width=1.3in]{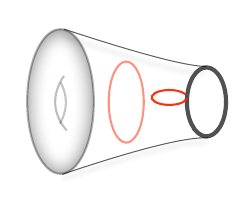}
	\end{aligned}
\end{equation}
where we slice the complement from the boundary of $L_1$ to the boundary of $L_2$ along the radial direction, corresponding to moving from the left boundary to the right boundary of the cobordism.
The longitude of $L_1$, if transported in a bulk, goes to the meridian of $L_2$, as shown in \eqref{cobor}.

To obtain a closed three-manifold, one can glue in solid tori to the boundaries of $L_1$ and $L_2$:\begin{equation}\label{hopftglueback}
	\begin{aligned}
		\includegraphics[width=2in]{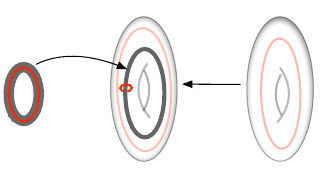}
	\end{aligned}
\end{equation}
In this figure,
matter circles are not linked, but in general matter circles have diverse ways to tangle and fuse with each other, which involve non-trivial phenomenon beyond the scope of this note. In \cite{Cheng:2024ybd}, we discuss the fusion of matter circles.
We show two examples in the following to show the diversity:
\begin{equation}\label{bifundcircle}
	\begin{aligned}
		\includegraphics[width=4in]{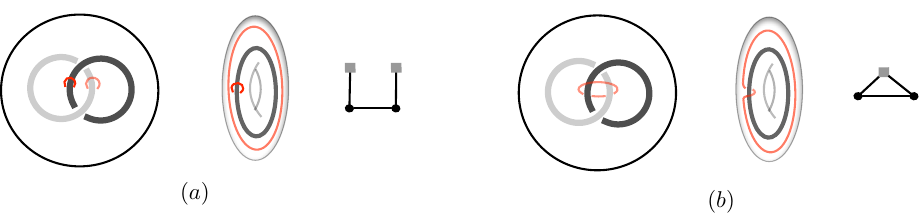}
	\end{aligned}
\end{equation}

It is possible to identify theories labeled by cobordisms.  
Let us first investigate the Hopf link in \eqref{hopftglueback}, which leads to a closed three-manifold with two matter circles: $M_3 ( \redcirc_1 ,\redcirc_2)=\big( S^3 \backslash( \bigcirc_1 \cup \bigcirc_2 ) \big) \sqcup_{f_1} (D^2 \times \redS)_1 \sqcup_{f_2} (D^2 \times \redS)_2 $.   Each solid torus with a red circle encodes a hypermultiplet, namely $T[D^2\times \redS] = 1\F +1\AF $, which reduces to a chiral multiple if $\AF$ is decoupled. In addition, the Hopf link carries a mixed CS level $k_{12}=+1$ and  the gauge group $U(1) \times U(1)$. If the Hopf link is given a  vanishing framing number and a gauge theory duality is applied, we get  \begin{equation}\label{Tsu2ST}
 	\begin{aligned}
 		\includegraphics[width=2in]{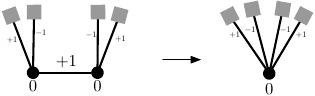}
 	\end{aligned}
 \end{equation}
The right graph represents the theory $U(1)_0 + 2\F+2\AF$, which is exactly the self-dual theory $T[SU(2)]$ discussed in \cite{Gaiotto:2008ak}. This example indicates that the Hopf link could engineer self-dual $T[SU(N)]$ theories if a number of M5-branes wrap on this three-manifold, although some problems may be involved for non-abelian theories.

\vspace{4mm}\noindent
\textbf{Gluing cobordisms.}
If $\bigcirc_1$ and $\bigcirc_2$ are not linked, then we have the complement $ S^3\backslash (\bigcirc_1 \cup \bigcirc_2) $ as follows,
\begin{equation}\label{twotori}
	\begin{aligned}
		\includegraphics[width=2in]{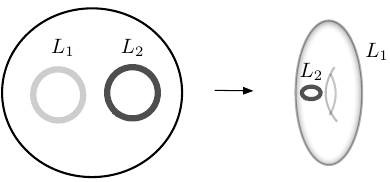}
	\end{aligned}
\end{equation}
where on the right graph we have taken the boundary of $L_1$ to the outside, and $L_2$ is still a torus hole.
We can glue two copies of the Hopf link complements in \eqref{cobor} to represent this complement:
\begin{equation}\label{cortori}
	\begin{aligned}
		\includegraphics[width=1.8in]{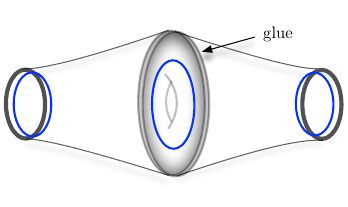}
	\end{aligned}
\end{equation}
where we glue the exterior boundary tori . Circles could  transport in this cobordism, that is, the meridian and longitude change as $\a \mapsto \a \mapsto \a$ and $\b \mapsto \b \mapsto \b$. 
If gluing two Hopf link complements $ S^3 \setminus L_1 \cup L'_2$ and $S^3\setminus L_2 \cup L_3$ by identifying $L'_2$ and $L_2$, one gets a larger link $S^3\setminus  L_1 \cup L_2 \cup L_3$, whose cobordism is
\begin{equation}\label{cobor2}
	\begin{aligned}
		\includegraphics[width=1.5in]{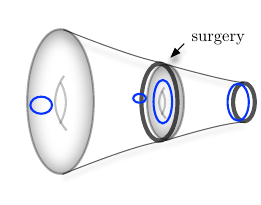}
	\end{aligned}
\end{equation}
where the torus switch is applied to glue the exterior boundary of $L_2$ and to the interior boundary of $L'_2$. The circles change: $\a \mapsto \a \xmapsto{\rm switch} \b \mapsto \b$ and $\b \mapsto \b \xmapsto{\rm switch} \a \mapsto \a$.

\section{ST-move as Rolfsen twist}\label{sec:STdrill}

3d gauge theories indicate that in the presence of chiral multiplets, one can apply the ST-moves. We have argued that the chiral multiplet can be realized by the OV Lagrangian branes. Then how to geometrically interpret these ST-moves? One may guess that it is also a Kirby move, but in fact Kirby moves work only on surgery circles rather than Lagrangian branes.
 In this section, we solve this puzzle by realizing ST-moves as Rolfen twists. It is known that Rolfsen twists are beyond Kirby moves, but give equivalent surgeries, and we have observed the Rolfsen twist playing an useful role in subsection \ref{imagecircs}, and in this section, we continue the discussion by translating equivalent twists to dualities.

\vspace{4mm}\noindent
\textbf{OV Lagrangian brane and the identical surgery.}
Recall that identical surgery is defined by the operation of cutting out a solid torus along any knot in the three-manifold and then filling in the same solid torus, so the three-manifold does not change.
A feature of the identical surgery is that the identical surgery should be given an infinite framing number by definition. For instance, the three-sphere $L(0,1)$ is given by an identical surgery along $\bigcirc_\inf$. For a given three-manifold $M_3 =S^3_f(K)$, one can freely perform an identical surgery along any knot ${K}' \in M_3$ which can either link or not link to the $K$; this however does not change the three-manifold itself.
 
In the presence of a matter circle $\redcirc$ in the three-manifold $M_3$, one can always perform an identical surgery $\redcirc_\inf$ on it neighborhood, since $\redcirc$ is just a loop in the three-manifold. The Ooguri-Vafa Lagrangian brane locates in the cotangent bundle and does not really intersect the three-manifold $M_3$, hence any operation of the base three-manifold does not affect the OV Lagrangian brane in the bundle.
Adding an OV Lagrangian brane can be represented by
\begin{equation}
	M_3( \redcirc) := M_3\backslash \bigcirc ~\cup_{\inf} ~\big( D^2 \times  \redcirc \big) \,
	\end{equation}
where the OV Lagrangian brane is introduced through the solid torus.

Since the identical circle $\redcirc_\inf$ has an infinite framing number, how to apply Kirby moves on it is a bit confusing.
Fortunately, the equivalent surgeries of the identical surgery can be described by the Rolfsen twist in \cite{Rolfesn:1984} to get rid of the infinity. For instance, this twist could describe $S^3_\inf$ as the lens space $L(\pm1,1)$, and hence the infinite framing number is gone. 

\vspace{4mm}\noindent
\textbf{Rolfsen twists.}
The Rolfsen twist does change linking numbers of surgical circles, and is basically an equivalent Dehn twist on the meridian by sending $\a' \mapsto \a'+ t\b'$ and $\b' \mapsto \b'$ as shown in \eqref{sphereimage}, and this twist turns $L(1,0)$ to $L(1,t)$. 
If we perform $t$ Rolfsen twists on a surgical circle $L$ with a framing number $k$, then the circles $L_i$  linking to ${L}$ are also twisted.  Assume that we have a link $L_1\cup L_2 \cup \cdots \cup L_m $ with linking  numbers
$L_i \cdot L_i :=k_i \,,~~L_i \cdot L_j :=k_{ij} $; then
applying the Rolfsen twist on $L$ gives a twisted circle $\tL$ with the framing number
\begin{align}
&
{\tL}\cdot{\tL} =	\tilde{k} = \frac{1}{t +\frac{1}{k} } \,, ~~ t \in \mathbb{Z} \,.
\end{align}
If  circles $L_i$ and $L_j$ are linked to $L$, then Rolfsen twists do not affect the linking numbers between them
$ {\tL}_i\cdot{\tL}_j=  {L}_i\cdot {L}_j = {k}_{ij} $, and $\tL_i \cdot {\tL} = {L_i} \cdot {L}$. The linking numbers between $L_i \cdot L_j$  become
\begin{equation}{\tL}_i \cdot {\tL}_j = L_i \cdot L_j + t (L_i \cdot L)(L_j \cdot L)  \,.\end{equation}
 Instead of reading these subtleties,
it is more convenient to use a graph to illustrate:
	\begin{equation}
\begin{aligned}\label{Rtwist}
	\includegraphics[width=2.5in]{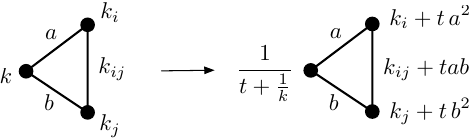}
\end{aligned}
		\end{equation}	
        where the left graph contains circle $L$ and $L_i$, and the right graph is given by Rolfsen twists, and contains circle $\tL$ and $\tL_i$.
This equivalent surgery gives a strong constraint on framing numbers, since framing numbers should be integers for gauge theories, and it is easy to check that Rolfsen twists do not fall into the class of Kirby moves and are particularly designed for identical surgery.

We can analyze images of matter circles, following the discussion in section \ref{imagecircs}.
For the lens space $L(p,q)$, the images of circles after Rolfsen twists are $\a \mapsto p (\a'+t \b') + q\b'$ and $\b \mapsto r \a' +s \b' $. For the case of $L(1,0)$, the 
identical surgery gives $\a \mapsto \a'$ and $\b \mapsto \b'$ with $r=0$. After the Rolfsen twist,
the image is  $\b \mapsto r \a' +(t r\pm1 )\b'$ and $\a \mapsto \a'+\b'$ which forces $t=1$ and the lens space becomes $L(1,1)$ with $r=1$, $\b \mapsto \a'\,,\a \mapsto \a'+\b'\,  $. Both have the same linking number with the image of $\a$, namely $ \text{Im}(\b) \cdot \text{Im}(\a) =  1$. In short, for the identical surgery, the matter circle is parallel to the surgical circle, while after the Rolfsen twist, the matter circle links to the surgical circle:
\begin{align}
   \b' \xrightarrow[]{\text{Rolfsen twist}\, t=1 } \a' \,.
\end{align}
Let us use this property to derive the ST-move.

\begin{figure}[h]
	\centering
		\includegraphics[width=4in]{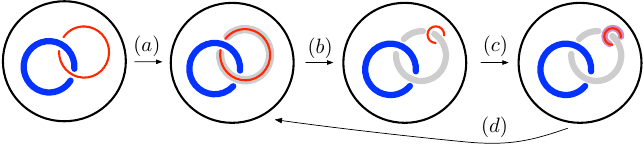}
		\caption{Using the identical surgery and Rolfsen twists, one can derive the $ST$-move.  }
		\label{fig:derivST}
\end{figure}

\vspace{4mm}\noindent
\textbf{A geometric derivation of the ST-move.}
Let us show a geometric derivation for the $ST$-move, which finally turns out to be a  Rolfsen twist supporting an OV Lagrangian brane. To prove this, we need to perform the drilling trick as shown in Figure \ref{fig:derivST}. The map $(a)$ is an identical surgery for the circle $\redcirc$,  since on the neighborhood of a loop in the three-manifold, one can always assume that there is an identical surgery around it. In the map $(b)$, this surgical loop is replaced by an equivalent surgery caused by a Rolfsen twist, which changes the framing number and the location of the circle $\redcirc$ from longitude to meridian. In the map $(c)$, one can repeat the identical surgery and the Rolfsen twist. One can observe that combined maps $(b)\circ(c)$ is the $ST$-move.

To introduce a charge $q$ chiral multiplet in $L(k,1)$, we need to change the associated linking number between matter circle and the gauge circle. Then the Rolfsen twist with $t=\pm1$ produces the following ST-move:
	\begin{equation}
		\begin{aligned}\label{effST}
	\includegraphics[width=2in]{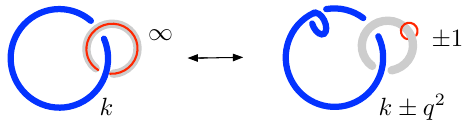}
\end{aligned}
		\end{equation}
		where we have assigned effective framing numbers. This is just the map $(b)\circ (c)$ in Figure \ref{fig:derivST}.
		 Linking numbers of circles match with that of $ST$-moves:
\begin{equation}\begin{aligned}\label{plumbSTk}
\includegraphics[width=1.2in]{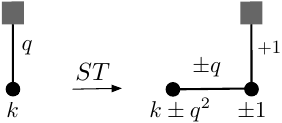}
\end{aligned}\end{equation}
which is equivalent to the graph in \eqref{bareSTq}. 
Ungauging the gauge group is equivalent to deleting the gauge nodes $\bullet_k$ and $\bullet_{k+q^2}$,
then the above graph reduces to 
\begin{equation}\label{STgraphex}
\begin{aligned}
	\begin{tikzpicture}
	\node(A) at (-0.7,0) {$\graybox $}; 
\draw[thick] (2,-0.5)--(2,0.5) ;
\draw[latex-latex] (0,0)--(1.2,0) node[midway,above]{\small{mirror}} ;
	  \node at (2,-0.5)[circle,fill,inner sep=2pt]{};
		\node(D) at (2,0.5) {$\graybox $}; 
		\node (E) at (2.2,0) { \tiny $\pm1$};
		\node at (2.3, -0.5){\tiny $ \pm1$};
		\end{tikzpicture}
	\end{aligned}
	\end{equation}
	which describes the basic mirror duality: $1\F \leftrightarrow U(1)_{\pm1} +1\F$.
Correspondingly, this ungauging deletes the blue circles in \eqref{effST} and  leaves a $L(1,0)$ with a circle $\redcirc$ on its longitude, which is dual to a $L(\pm 1,1)$ with a $\redcirc$ on its meridian. Using graphs, this mirror duality (ST-duality) is represented as
\begin{equation}\label{mirrortrialitycirc}
	\begin{aligned}
	\includegraphics[width=1.5in]{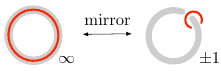}
	\end{aligned}
	\end{equation}
This $ST$-move is highly nontrivial, as it means that even if the matter circle $\redcirc$ is present, the Rolfsen twist is still a well defined homeomorphism.

\begin{figure}[h]
	\centering
		\includegraphics[width=2.5in]{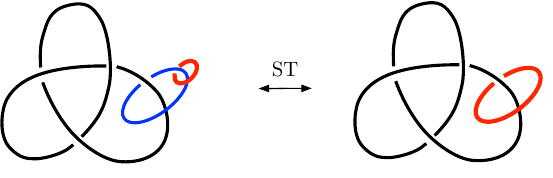}
		\caption{This three-manifold is given by a trefoil.  }
		\label{fig:STex1}
\end{figure}

 \vspace{4mm}\noindent
\textbf{Knot examples.} 
Recall that if a three-manifold is given by the surgery along a knot, then this knot can be transformed into a link by Kirby moves.
Even if a red circle $\redcirc$ is linked to a knot $K$, one can still apply $ST$-moves. An example is shown in Figure \ref{fig:STex1}. 
Trefoil is often used as a typical example in textbooks, e.g. \cite{Prasolovbook,Rolfesn:1984,Rolfsenbook}, because it can be turned into a link by Kirby moves. Here we apply the $ST$-move on the circle $\redcirc_\inf$: 
\begin{equation}
	\begin{aligned}
		\includegraphics[width=2.2in]{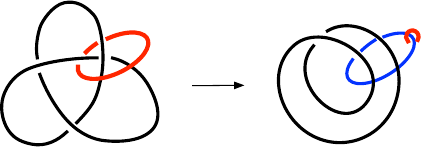}
	\end{aligned}
\end{equation}
where a charge-two chiral multiplet is shown on the left graph, and on the right graph, a surgical link is obtained through $ST$-move.
Note that matter circles could tangle with trefoils in any complicated way, and in general the red circle can be any element in the fundamental group of the knot complement $\pi_1(S^3\backslash K )$.

By now, we have considered the cases that matter circles are linked to surgical links or knots. The case of matter circles linking to each other are beyond the scope of this paper. It is easy to check that the naive application of $ST$-moves in this case leads to wrong results, which implies that linked matter circles are highly nontrivial and require other methods to address. For instance, we cannot naively separate a Hopf link of Lagrangian branes by $ST$-moves:
	\begin{equation}
	\begin{aligned}\label{effST2}
		\includegraphics[width=4.5in]{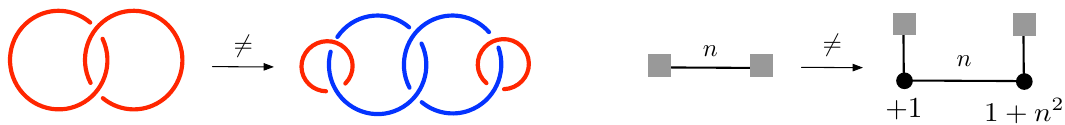}
	\end{aligned}
\end{equation}
It is unclear if there exists a new move that  we can use to separate linked matter circles.

The linked matter circles are indeed exist and can be produced through gluing maps. Given a lens spaces $L(k,1)$ with two Lagrangian branes on its two endpoints, respectively; the images of these two circles are $r_1 \a' +s_1 \b'$ and $r_2 \a'+s_2 \b'$. To preserve $SL(2,\mathbb{Z})$, we define integers $m_1=(r_1+1)/k$ and $m_2=(r_2+1)/k$, and the linking number of these two matter circles can be computed as $2 m_1 m_2 k - (m_1+m_2)$. One can find a solution that $m_1=1, m_2=0$ leads to a Hopf link of matter circles. This solution corresponds to the exceptional case of 3d brane webs that we discussed before. This example shows that linked matter circles can be generated by gluing maps and Lagrangian branes. 
For introducing a linked matter in a generic three-manifold, we can draw a Hopf link in the three-manifold, and then consider two identical surgeries supporting OV Lagrangian branes on this Hopf link.

\paragraph{Mass and FI parameters revisited.}\label{sec:FImassrev}

We claim that the FI parameter that we  mentioned in Section \ref{sec:matterdefect} can be determined by ST-duality. The ST-duality in \eqref{STgraphex} is between a chiral multiplet and a theory with $U(1)_{\pm1}$ and a chiral multiplet. This is a basic mirror symmetry that the flavor symmetry of $U(1)_F$ of the free chiral multiplet is dual to the topological symmetry $U(1)_T$ of the $U(1)_{\pm1}$ theory. Thus, mass parameter should be dual to the FI parameter $ m \leftrightarrow \xi$. 
This indicates that the
length of the red circle for the identical surgery in \eqref{mirrortrialitycirc} is the mass parameter $m$, and on the right graph the red circle winding the meridian of the surgery circle should give the FI parameter $\xi$. Since these two parameters are associated to the length of the same red circle, we have $m=\xi$. 

There is an easy interpretation from open topological strings and geometric transition. We can perform geometric transition on $\L_\redcirc$, such that the boundary of the M2-brane tube $\redcirc \times I$ on $\L_\redcirc$ closes up and the M2-brane becomes a disc. Locally, this is  open topological strings  on the toric geometry $\mathbb{C}^3$.
\begin{align}
\begin{split}
\includegraphics[width=1in]{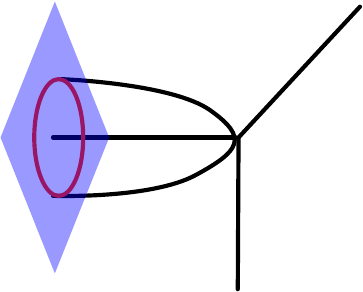}
\end{split}
\end{align}
where the blue rectangle is the D3-brane coming from the closed three-manifold, and the $\mathbb{C}^3$ comes from the geometric transition of $\L_{\redcirc}$.
Vortex partition function shows that this M2-brane disc gives  the partition function of $U(1)_1+1 \F$ for which the length of its boundary $\redcirc$ is the FI parameter, and the chiral multiplet is massless. This theory just describes the right graph in \eqref{mirrortrialitycirc}. 

Note that although we put a single M5-brane on $\L_\redcirc$, the geometric transition \cite{Gopakumar:1998ki} can still be applied. The "magic" here is that the M2-brane disc has a radius equal to the height of the original M2-brane tube, namely the height $I$ is equal to the length of $\redcirc$. This agrees with our previous conclusion that the $I$ should be the mass parameter, and verifies the dictionary between the three-manifolds and 3d brane webs, at least for lens spaces. In addition, the geometric transition/conifold transition can be viewed as integrating out symmetry groups on boundaries, which can be done by adding an integral to the path integrals of partition functions. If integrating out on the $\L_\redcirc$ side, the flavor symmetry is gone, and if integrating out on the side of the closed three-manifold, the gauge group is gone. 
Let us further discuss the latter case.
The length of the $\redcirc_\inf$ is a bit confusing when it does not link to any gauge circle, since in this case the $\redcirc_\inf$ is shrinkable to a point on the closed three-manifold, while the boundary ending on the $\L_\redcirc$ is not shrinkable, so the M2-brane tube in this case also reduces to a disc but with its tip ending on the closed three-manifold, and it is not charged under a gauge group, so this M2-brane gives a free chiral multiplet that describes the left graph in \eqref{mirrortrialitycirc}. 

To summarize, the meridian of the surgical circle $\bigcirc_k$ with $k \neq \inf$ is the FI parameter, and the longitude of the identical surgical circle $\redcirc_\inf$ supporting a Lagrangian brane $\L_{\redcirc}$ is the mass parameter.

\acknowledgments
I would like to thank Satoshi Nawata for helpful discussions and the hospitality of UESTC and BIMSA where parts of this work were done. 
The work of SC is supported by the NSFC Grant No.12050410234 and the NSFC Grant No.12305078, and the start-up grant of SIMIS.

\appendix

\section{Lens spaces}
Lens spaces are defined as orbifolds of the three-sphere: $L(k,1) = S^3/\mathbb{Z}_k$. With complex coordinates, the three-sphere is writen as $|z_1|^2+|z_2|^2 = r^2$, and the orbifold action $\Z_k$ is $(z_1, z_2) \mapsto (e^{ \frac{2 \pi \, i}{k}} z_1, e^{ \frac{2 \pi \, i}{k}} z_2  ) $. The homology of the lens space is a torsion $H_1(L(k,1)) = \Z_k$. Some lens spaces are well known, such as $L(2,1) =S^3/\Z_2 = \mathbb{RP}^3$.  
Lens spaces have many equivalences, for example, $L(p,q) = L(p, q+np)=L(-p,q)$, which derives $L(k,1)= L (\frac{k}{nk+1},1)$ caused by Rolfsen twists. 
The cotangent bundle $T^*L(k,1)$ of lens spaces is a noncompact Calabi-Yau three-manifold, and the lens space as the base is a
also Lagrangian submanifold.  As examples of Seifert manifolds, lens spaces are circle bundles over a two-sphere: $\mathcal{O}(-k) \hookrightarrow L(k,1) \xrightarrow{\pi} {\mathbb{P}^1} $.

In terms of surgeries, the lens space is obtained by surgery along an unknot, namely $L(k,1) :=S^3( \bigcirc_k)= (S^3- N (\bigcirc)) \cup_k (D^2 \times S^1)$ where $N(\bigcirc)$ is the neighborhood of the surgery circle. The plumbing graph is a shorthand for surgical links by
\begin{equation}
	\begin{aligned}
\includegraphics[width=3.5in]{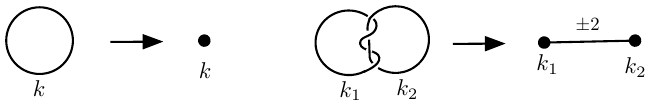}
	\end{aligned}
\end{equation}
where linking numbers are assigned on the line segment. Depending on orientations, linking numbers can be given opposite signs. 

The framing number is defined as the linking number between the surgical circle $L$ and its slightly deformed circle $L'$, which is $L_i\cdot L_i:= L_i \cdot L'_i =k_i$. The linking number is also a half of the crossing number between two surgery circles: $L_i\cdot L_j=k_{ij}:=\frac{1}{2} \times \text{crossing number} $. 

The framing number can also be read from the gluing maps between two solid tori.
We can denote the meridian by $\a$ and the longitude by $\b$ for a solid torus, and the other solid torus is the unknot complement. The gluing map as an element of the mapping class group $SL(2,\mathbb{Z})$ maps the circles on the boundary of the solid torus to the unknot complement:
\begin{align}\label{gluingmaplen}
	\begin{bmatrix}
		f({\a})  \\
		f({\b}) 
	\end{bmatrix}=
	\begin{bmatrix}
		k & 1\\
		-1  &0 
	\end{bmatrix} \cdot
	\begin{bmatrix}
		{\a}  \\
		{\b} 
	\end{bmatrix}
	\,,
\end{align}
which encodes the framing number $k$. 

The  matter circle of charge $q$ is the connected sum of $q$ copies of the charge-$1$ matter circle, given by ${\b}_q= q \, \b$.
The linking numbers between the gauge circles and the matter circles  are invariant under the gluing maps: $ \a \cdot \b_q =f(\a) \cdot f({\b_q})=q $ with $\a \cdot \a =0\,, ~ \b \cdot \b =0\,,~ \a \cdot \b =1$.

\bibliographystyle{JHEP}

\end{document}